\documentclass[12pt]{article}
\usepackage[margin=0.5in]{geometry}
\usepackage{fullpage}
\DeclareUnicodeCharacter{FFFD}{\textquestiondown}
\DeclareUnicodeCharacter{03B2}{\ensuremath{\beta}}
\usepackage[utf8]{inputenc} 
\usepackage{booktabs} 
\usepackage{mathptmx}
\usepackage[numbers,sort]{natbib}  
\usepackage{amsmath}  
\usepackage{lipsum}
\usepackage{multirow}
\usepackage{amssymb}
\usepackage[T1]{fontenc}
\usepackage{bm}
\usepackage{xcolor}
\usepackage{ulem}
\usepackage{graphicx}
\usepackage{caption}
\usepackage{comment}
\usepackage{pdfpages}
\usepackage{tabularx}
\usepackage{abstract}
\usepackage[colorlinks=true,linkcolor=re,citecolor=blue,urlcolor=cyan]{hyperref}
	\title{\large \textbf{Exploring the Thermodynamic, Elastic, and Optical properties of LaRh$_2$X$_2$ (X = Al, Ga, In) low T$_c$ Superconductors through First-Principles Calculations}}
	
		\author{ 
			\small 
					\begin{tabular}{c}
					\textsuperscript{1}Md Hasan Shahriar Rifat,
					\textsuperscript{1*}Mirza Humaun Kabir Rubel,
					\textsuperscript{1*}M. M. Rahaman\\
					\textsuperscript{1}Md Borhan Uddin,
					\textsuperscript{2}Apon Kumar Datta,
					\textsuperscript{3} Jubair Hossan Abir, \\ [6pt]
					\textsuperscript{1}Department of Materials Science and Engineering, University\\ of Rajshahi, Rajshahi 6205, Bangladesh \\
					\textsuperscript{2}Department of Electrical and Electronic Engineering, Mymensingh \\
					Engineering College, Mymensingh 2200, Bangladesh   \\
					 \textsuperscript{3}Department of Physics, University of Rajshahi, Rajshahi 6205, Bangladesh, \\ 
					Corresponding author’s e-mail:\textsuperscript{*} \href{mhk_mse@ru.ac.bd }{\uline{mhk\_mse@ru.ac.bd }},
					\href{mijan\_mse@ru.ac.bd }{\uline{mijan\_mse@ru.ac.bd }}
				\end{tabular}%
			}
	
	\date{} 

\begin{document}
\maketitle 
	\begin{abstract}
   		LaRh\textsubscript{2}X\textsubscript{2} (X = Al, Ga, In) possesses a tetragonal layered structure and belongs to the family of ThCr\textsubscript{2}Si\textsubscript{2}-type low-\textit{T\textsubscript{c}} superconductors. For the first time, we have investigated the structural, mechanical, elastic, electronic, vibrational, thermophysical, and optical properties using first-principles calculations based on density functional theory (DFT) implemented in the CASTEP code. The calculated structural parameters show good agreement with the available experimental results. Analysis of the Born stability criteria and Mulliken population confirms the mechanical stability of the materials. The values of Poisson's and Pugh’s ratios indicate the ductile behavior of these superconductors. Additionally, the low Debye temperature and melting point values, as inferred from the elastic constants, suggest the soft nature of the materials. The electronic band structure and density of states (DOS) demonstrate metallic behavior. The partial DOS indicates a strong contribution from Rh-4\textit{d} and Rh-4\textit{p} states, whereas Rh-4\textit{s} states contribute minimally. Electronic charge density mapping and Mulliken population analysis reveal a mixed bonding nature: covalent, ionic, and metallic. The calculated Fermi surfaces show both hole-like and electron-like sheets, indicating possible multi-band superconductivity. Vickers hardness calculations further confirm the soft nature of the studied materials. Phonon dispersion curves suggest vibrational stability at the $\Gamma$ point for LaRh\textsubscript{2}Al\textsubscript{2} and LaRh\textsubscript{2}Ga\textsubscript{2}, while LaRh\textsubscript{2}In\textsubscript{2} shows slight instability. The optical functions are extensively analyzed: the refractive index values indicate potential for high-density optical data storage at ambient conditions. Optical absorption is notably strong in the high-energy ultraviolet region, making these materials suitable for solar cell applications. Finally, the calculated electron-phonon coupling constant $\lambda = 0.56$ for LaRh\textsubscript{2}Ga\textsubscript{2} implies that it is a weakly coupled low-\textit{T\textsubscript{c}} superconductor.
	\end{abstract}
	\textbf{keywords:} Tetragonal layered structure; Elastic properties; Optoelectronic properties, Phonon properties; Thermodynamic and Superconducting properties; DFT computations.

	\section{INTRODUCTION}
	Superconductivity is a quantum mechanical phenomenon observed in certain materials at cryogenic temperatures, and represents a cornerstone of modern applied science. The advent of superconductivity has since sparked a century of extensive research, unveiling a myriad of materials and mechanisms underlying this extraordinary behavior. From the foundational theories established by Bardeen, Cooper, and Schrieffer (BCS theory) to the discovery of high-temperature superconductors, the field has evolved with profound implications for both fundamental science and technological innovation \cite{c1}. Superconductors have the potential to revolutionize technology by enabling faster and more efficient systems. However, their widespread application is limited by the fact that most superconductors operate only at extremely low temperatures, requiring costly and complex cooling systems. This challenge has driven researchers to focus on developing materials that exhibit superconductivity at room temperature \cite{c2,c3,c4,c5,c6,c7}, 
	Rubel \textit{et al.} and other scientists have nevertheless recently investigated the Bi-based superconductive \( \mathrm{AA'}_3\mathrm{B}_4\mathrm{O}_{12} \)--type perovskite structure \cite{c8,c9}, which is also referred to as double perovskite according to their unit cell volume. Additionally, a number of theoretical investigations are being conducted on these Bi-oxide double and single perovskites \cite{c10,c11,c12}. In many years, the field of condensed matter has effectively used DFT-based first-principle computations. In recent years, it has emerged as a significant and prospective method for forecasting a wide range of material physical traits \cite{c13,c14,c15}.
	
	Interestingly, ThCr\textsubscript{2}Si\textsubscript{2}, i.e., AT\textsubscript{2}X\textsubscript{2}-type (where A = lanthanide or alkaline earth elements; T = transition metals; X = P, Se, Si, Ge, or As) ternary intermetallic compounds, were first introduced in 1965\cite{c16} by Ban and Sikirica, have garnered significant attention for their rich physical properties and potential applications in superconductivity field. These body-centered tetragonal compounds, often referred to as the "122 family," exhibit fascinating chemical and physical phenomena that make them a focal point of condensed matter research \cite{c17}.Among them, several iron-free ThCr\textsubscript{2}Si\textsubscript{2}-type compounds, such as SrNi\textsubscript{2}As\textsubscript{2}, LaRu\textsubscript{2}P\textsubscript{2}, and BaRh\textsubscript{2}P\textsubscript{2}, have already demonstrated superconducting behavior, albeit with very low transition temperatures \cite{c11,c12,c13,c14}. These materials offer a promising platform for exploring novel phenomena and advancing the development of new superconductors. La-incorporated LaRh\textsubscript{2}X\textsubscript{2} (X = Al, Ga, In) compounds are usually nonmagnetic/diamagnetic and promising superconductors with a low superconducting transition temperature \( T_c \) of 3.7~K \cite{c22}, exhibiting intriguing physical, optical, and electronic properties. Despite prior experimental investigations, a comprehensive understanding of their fundamental properties remains incomplete. Critical aspects such as elastic, mechanical, and bonding characteristics, as well as phonon dispersion and optical behavior, require further exploration. These properties are essential for evaluating the stability and anisotropy of the materials, which might play a pivotal role in their potential applications \cite{c17,c18,c19,c20,c21}. In particular, understanding the elastic, mechanical, optoelectronic, and phonon dispersion properties is crucial for assessing the structural and physical attributes of these compounds.
	
	Therefore, the contents of this paper are organized into three sections to explore unrevealed interesting physical characteristics for the first time. The first section outlines the computational methods employed in this investigation, detailing the techniques and frameworks used to analyze the properties of the studied compounds. The second section focuses on the analysis of elastic, optoelectronic, vibrational, bonding, thermodynamic, and superconducting properties to evaluate the underlying mechanisms and related physical behaviors. Finally, the third section summarizes the key findings and presents the conclusions drawn from this study, highlighting their significance and potential implications for future applications and research of LaRh\textsubscript{2}X\textsubscript{2} (X = Al, Ga, In).

	\section{Computational Details }
	
	First-principles calculations were performed within the context of Density Functional Theory (DFT)~\cite{c23} utilizing the Materials Studio Simulation Software and the Cambridge Serial Total Energy Package (CASTEP)~\cite{c24}. In the Perdew-Burke-Ernzerhof (PBE) function, the Generalized Gradient Approximation (GGA) was utilized to incorporate electronic exchange-correlation energy terms in the total energy calculations~\cite{c25}. The energy of the Coulomb potential resulting from the interaction between ion cores and valence electrons was modeled using the Vanderbilt-type Ultra-soft Pseudopotential~\cite{c26}. This approach significantly reduces computational time with minimal loss in accuracy. The BFGS (Broyden-Fletcher-Goldfarb-Shanno) algorithm was applied as a minimization strategy to determine the ground state energy of the material system~\cite{c27}. The electronic configurations considered were:
	La [5s\textsuperscript{2} 5p\textsuperscript{6} 5d\textsuperscript{1} 6s\textsuperscript{2}],
	Rh [4s\textsuperscript{2} 4p\textsuperscript{6} 4d\textsuperscript{8} 5s\textsuperscript{1}],
	Al [3s\textsuperscript{2} 3p\textsuperscript{1}],
	Ga [3d\textsuperscript{10} 4s\textsuperscript{2} 4p\textsuperscript{1}], and
	In [4d\textsuperscript{10} 5s\textsuperscript{2} 5p\textsuperscript{1}]. A plane wave cutoff energy of 600 eV was used, and Brillouin zone sampling was performed using a Monkhorst-Pack grid of 9$\times$9$\times$4. Convergence thresholds were set to:
	10\textsuperscript{-5} eV/atom for total energy,
	0.03 eV/\AA{} for maximum force,
	0.05 GPa for maximum stress, and
	10\textsuperscript{-3} \AA{} for maximum atomic displacement. These parameters ensured convergence to a reliable ground state configuration for all studied compounds.
	
	\section{RESULTS AND DISCUSSION}
	\subsection{Structural properties}
	The compounds LaRh$_2$Al$_2$, LaRh$_2$Ga$_2$, and LaRh$_2$In$_2$ crystallize in a tetragonal structure with the space group \textit{I4/mmm} (No.~129)\cite{c22}. Diverse structural configurations of these investigated materials are visualized using VESTA software, as shown in Fig.~1(a--c). In these compounds, La atoms occupy the 2$c$ Wyckoff site with coordinates $(-0.5,\, 0.0,\, -0.24129)$, while Rh atoms are distributed between 2$c$ sites at $(0.0,\, 0.5,\, 0.6261)$ and 2$a$ at $(0.0,\, 0.0,\, 0.0)$. Al, Ga, or In atoms occupy 2$c$ and 2$b$ positions, exemplified by Ga coordinates of $(0.0,\, 0.5,\,-0.12245)$ and $(-0.5,\, 0.5,\, 0.5)$ in Table 1. Equilibrium crystal structures of LaRh$_2$Al$_2$, LaRh$_2$Ga$_2$, and LaRh$_2$In$_2$ are determined by minimizing their total energy. Geometry optimization of LaRh$_2$X$_2$ compounds is performed at zero temperature and pressure to reveal their ground-state properties. The lattice parameters and cell volume are obtained through geometry optimization, as presented in Table~2.which shows a close agreement with the experimental values. This strong agreement further validates the accuracy and reliability of this study. According to Tables~2 and~6, Rh--Al, Rh--Ga, and Rh--In bond lengths increase for LaRh$_2$Al$_2$, LaRh$_2$Ga$_2$, and LaRh$_2$In$_2$ structures, which also correlates with the corresponding increase in cell volume.\\
	\begin{figure}[ht]
		\centering
		\includegraphics[width=0.98\textwidth]{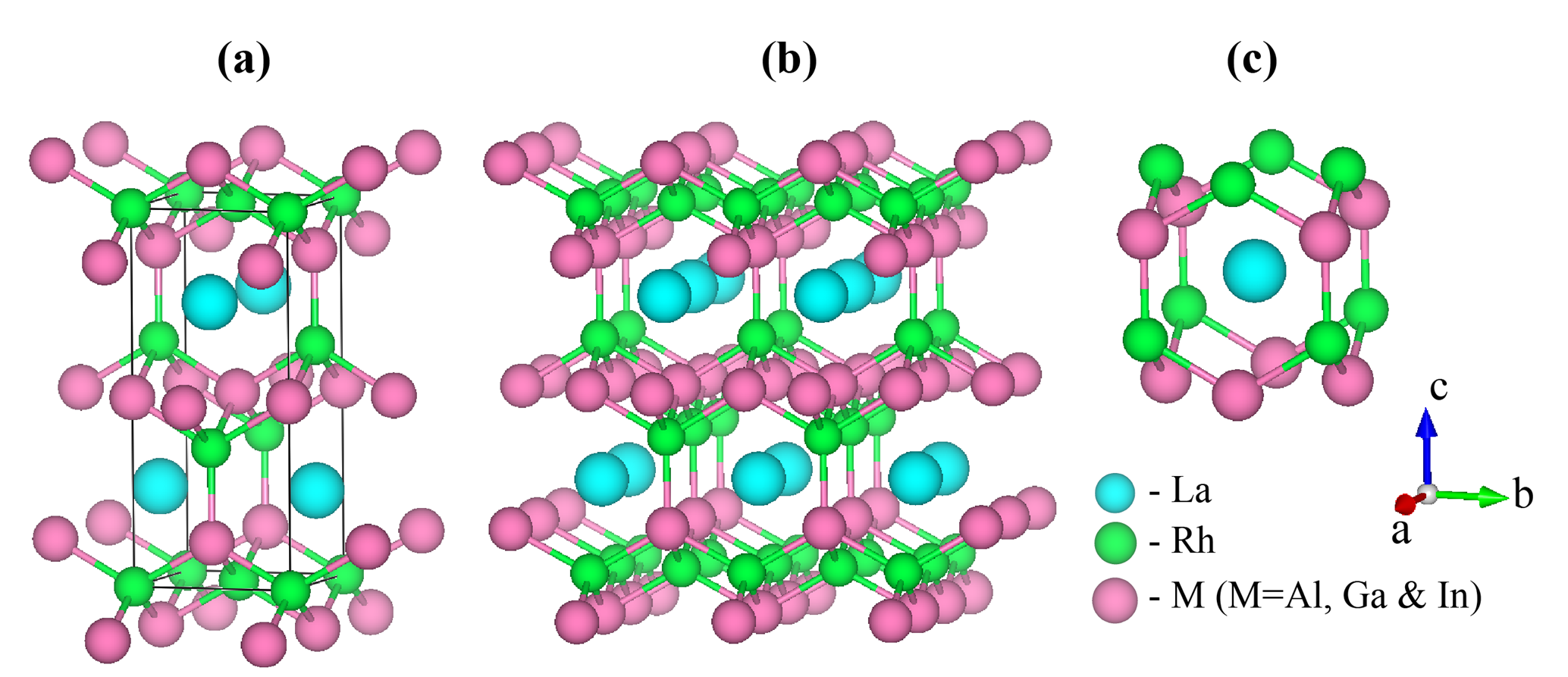}
		\caption{\label{Crystal structure}(a) Unit cell, (b) supercell ($2\times2\times1$) of LaRh2X2 and (c) visual representation of covered by LaRh\textsubscript{2}X\textsubscript{2} (X = Al, Ga, In) structure. }
	\end{figure}
	
	\begin{table}[htbp]
		\centering
		\caption{Wyckoff positions of the different atomic species in LaRh$_2$X$_2$ crystal structures.}
		\begin{tabular}{|c|c|c|c|c|c|}
			\hline
			\textbf{Compound} & \textbf{Wyckoff} & \textbf{Atom} & \textbf{x} & \textbf{y} & \textbf{z} \\
			\hline
			LaRh$_2$Al$_2$ & 2a & Rh & 0   & 0   & 0     \\
			& 2b & Ga & 0.5 & 0.5 & 0.5   \\
			& 2c & La & 0   & 0.5 & 0.248 \\
			& 2c & Rh & 0.5 & 0   & 0.365 \\
			& 2c & Al & 0.5 & 0   & 0.135 \\
			\hline
			LaRh$_2$Ga$_2$ & 2c & La & 0   & 0.5 & 0.243 \\
			& 2c & Rh & 0.5 & 0   & 0.371 \\
			& 2c & Ga & 0.5 & 0   & 0.121 \\
			\hline
			LaRh$_2$In$_2$ & 2c & La & 0   & 0.5 & 0.236 \\
			& 2c & Rh & 0.5 & 0   & 0.369 \\
			& 2c & In & 0.5 & 0   & 0.124 \\
			\hline
		\end{tabular}
	\end{table}

	\begin{table}[htbp]
		\centering
		\caption{Lattice parameters, cell volume, and formation energy of LaRh$_2$X$_2$ structure.}
		\label{tab:lattice}
		\begin{tabular}{lcccccc}
			\toprule
			\textbf{Compounds} & \textbf{Total Energy} & \textbf{Cal./Exp. $a$ (\AA)} & \textbf{Cal./Exp. $c$ (\AA)} & \textbf{$V$ (\AA$^3$)} & \textbf{$\Delta E_f$ (eV/atom)} \\
			\midrule
			LaRh$_2$Al$_2$ & -4387.9   & 4.35           & 10.14          & 192.29 & -6.19 \\
			LaRh$_2$Ga$_2$ & -12372.74 & 4.38/4.3423    & 10.12/9.9166   & 194.84 & -5.94 \\
			LaRh$_2$In$_2$ & -10401.38 & 4.61           & 10.84          & 229.91 & -5.39 \\
			\bottomrule
		\end{tabular}
	\end{table}

	Formation energy, indicating the energy necessary to disrupt the bonds between various atoms in a crystal structure, was calculated for LaRh$_2$X$_2$ materials using the following equation:
	
	\begin{equation}
		\Delta E_f (\mathrm{LaRh}_2\mathrm{X}_2) = \frac{E_\mathrm{total} - (E_\mathrm{La} + 2E_\mathrm{Rh} + 2E_\mathrm{X})}{10}
	\end{equation}
	Here, $E_\mathrm{tot}$ is the system's total energy, while $E_\mathrm{La}$, $E_\mathrm{Rh}$, and $E_\mathrm{X}$ correspond to the energies of La, Rh, and X (Al, Ga, In) atoms in their respective stable crystal configurations. Table~2 depicts the formation energy of the investigated structures, and it is evident that all the structures have negative formation values, confirming their thermodynamic stability and suitability for the fabrication process.
	
	\subsection{Mechanical Properties}
	\subsubsection{Elastic constant}
	Solids' elastic characteristics are linked to atomic bonding and cohesive energy. By analyzing elastic constants and moduli, the behavior of solids under stress can also be determined. Mechanical and other physical characteristics of crystalline compounds are associated with elastic properties. The response of a material to applied stress, or mechanical stability, can be ascertained in part by using elastic constants. The tetragonal LaRh$_2$X$_2$ (X = Al, Ga, In) structures possess six distinct elastic constants: $C_{11}$, $C_{12}$, $C_{13}$, $C_{33}$, $C_{44}$, and $C_{66}$~\cite{c28}, and their calculated values are presented in Table~3. According to Born-Huang conditions, the necessary and sufficient criteria for mechanical stability of a tetragonal system are as follows Eqs. 2.
	
	\begin{equation}
		\left\{
		\begin{aligned}
			& C_{11} > 0,\; C_{33} > 0,\; C_{44} > 0,\; C_{66} > 0 \\
			& C_{11} - C_{12} > 0,\; C_{11} + C_{33} - 2C_{13} > 0 \\
			& 2(C_{11} + C_{12}) + C_{33} + 4C_{13} > 0
		\end{aligned}
		\right\}.
		\tag{2}
	\end{equation}
	LaRh$_2$X$_2$ compounds meet the mechanical stability requirements by exhibiting positive elastic constants, indicating their superior mechanical stability. The resistance to linear compression along [100] and [001] is regarded as the elastic constants $C_{11}$ and $C_{33}$, respectively. The observation that $C_{33}$ is greater than $C_{11}$ for LaRh$_2$Ga$_2$ and LaRh$_2$In$_2$ suggests that the bonding strengths along the [001] direction are stronger than that along the [100] direction. The resistance to shear deformation in response to a tangential stress applied to the (100) plane in the compounds [010] direction is represented by the elastic constant $C_{44}$. For all of the three investigated compounds, $C_{44}$ is significantly less than $C_{11}$ and $C_{33}$, suggesting that shear can deform the compound more readily than unidirectional stress can along any of the three crystallographic orientations. The Voigt–Reuss–Hill averaging techniques can be used to evaluate the polycrystalline elastic modulus for the tetragonal crystal, including the bulk modulus ($B_\mathrm{VRH}$), the shear modulus ($G_\mathrm{VRH}$), Young’s modulus ($E$), Poisson ratio ($\nu$), and elastic anisotropic factor ($A^U$)~\cite{c29}.
	\begin{equation}
		B_v = \frac{2}{9}\left( C_{11} + C_{12} + 2C_{13} + \frac{C_{33}}{2} \right)
		\tag{2}
	\end{equation}
	
	\begin{equation}
		B_R = \frac{C^2}{M}
		\tag{3}
	\end{equation}
	
	\noindent
	Where, $C^2 = (C_{11} + C_{12})C_{33} - 2C_{13}^2$\\
	And $M = C_{11} + C_{12} + 2C_{33} - 4C_{13}$
	
	\begin{equation}
		G_V = \frac{M + 3C_{11} - 3C_{12} + 12C_{44} + 6C_{66}}{30}
		\tag{4}
	\end{equation}
	\begin{equation}
		G_R = \frac{15}{\left(\frac{18B_V}{C^2} + \frac{6}{C_{11} - C_{12}} + \frac{6}{C_{44}} + \frac{3}{C_{66}}\right)}
		\tag{5}
	\end{equation}
	
	The Hill took an arithmetic mean of $B$ and $G$, which can be evaluated by the following equations,
	\begin{equation}
		B_H = \frac{B_v + B_R}{2}
		\tag{6}
	\end{equation}
	\begin{equation}
		G_H = \frac{G_v + G_R}{2}
		\tag{7}
	\end{equation}
	
	Moreover, the values of Young’s modulus $E$ and Poisson’s ratio $\nu$ can also be calculated by using the following expressions,
	\begin{equation}
		E = \frac{9B_H G_H}{3B_H + G_H}
		\tag{8}
	\end{equation}
	
	\begin{equation}
		\nu = \frac{3B_H - E}{6B_H}
		\tag{9}
	\end{equation}

\begin{table}[htbp]
	\centering
	\caption{Elastic constants $C_{ij}$ (GPa) of LaRh$_2$Al$_2$, LaRh$_2$Ga$_2$ and LaRh$_2$In$_2$ superconductors.}
	\label{tab:elastic_constants}
	\begin{tabular}{lccccccc}
		\toprule
		\textbf{Compound} & $C_{11}$ & $C_{12}$ & $C_{13}$ & $C_{33}$ & $C_{44}$ & $C_{66}$ & \textbf{Ref.} \\
		\midrule
		LaRh$_2$Al$_2$ & 161.24 & 80.23 & 90.04 & 155.17 & 33.66 & 35.59 & This \\
		LaRh$_2$Ga$_2$ & 163.33 & 87.33 & 97.11 & 164.39 & 32.47 & 29.16 & This \\
		LaRh$_2$In$_2$ & 112.17 & 82.91 & 87.58 & 121.64 & 5.53  & 21.74 & This \\
		\bottomrule
	\end{tabular}
\end{table}

	The subscripts V, R, and H in all the above equations indicate to Voigt, Reuss, and Hill, respectively. Table~4 shows the calculated values for the polycrystalline elastic modulus of the tetragonal crystal as well.
	The Pugh ratio ($B/G$) is another physical property that indicates the nature of a material, whether it is brittle or ductile. In general terms, a compound is considered ductile if its $B/G$ ratio is more than 1.75; if not, it is considered as fragile. All investigated superconductors exhibit a Pugh's ratio greater than 1.75~\cite{c30}, indicating ductility in each of the compounds as shown in Fig.~2(a). Notably, ductility and brittleness of a material can also be evaluated by using Poisson's ratio ($\nu$). If the value of $\nu$ is less than 0.26, the material is brittle; otherwise, it is ductile. All the three superconductors with greater values of 0.26 are therefore confirmed to display ductile behavior and the ductility behaviour is maximum for LaRh$_2$In$_2$ as depicted in Fig.~2(b). The Young's modulus ($E$) values for LaRh$_2$X$_2$ (X = Al, Ga, In) are illustrated in Table~4, where LaRh$_2$Al$_2$ exhibits the highest rigidity among the three compounds with a value of 94.89~GPa. Conversely, LaRh$_2$In$_2$ is comparatively less rigid with a value of 31.22~GPa.
	
	\begin{table}[ht]
		\centering
		\caption{The calculated values of bulk modulus, $B$ (GPa) and shear modulus, $G$ (GPa) by Voigt-Reuss-Hill method, along with Pugh’s ratio ($B/G$), Young’s modulus, $E$ (GPa), Poisson’s ratio ($\nu$) and universal elastic anisotropy ($A^U$).}
		\label{tab:polycrystalline_modulus}
		\begin{tabular}{lccc}
			\toprule
			\textbf{Compound} & \textbf{LaRh$_2$Al$_2$} & \textbf{LaRh$_2$Ga$_2$} & \textbf{LaRh$_2$In$_2$} \\
			\midrule
			$B_V$      & 110.91 & 117.12 & 95.79 \\
			$B_R$      & 110.90 & 116.99 & 95.28 \\
			$B$        & 110.91 & 117.05 & 95.53 \\
			$G_V$      & 35.07  & 32.78  & 12.42 \\
			$G_R$      & 34.83  & 32.54  & 9.18  \\
			$G$        & 34.95  & 32.66  & 10.80 \\
			$B/G$      & 3.17   & 3.58   & 8.84  \\
			$E$        & 94.89  & 89.65  & 31.22 \\
			$\nu$      & 0.35   & 0.37   & 0.44  \\
			$A^U$      & 0.034  & 0.038  & 1.77  \\
			\bottomrule
		\end{tabular}
	\end{table}
	\begin{figure}[ht]
		\centering
		\includegraphics[width=0.5\textwidth]{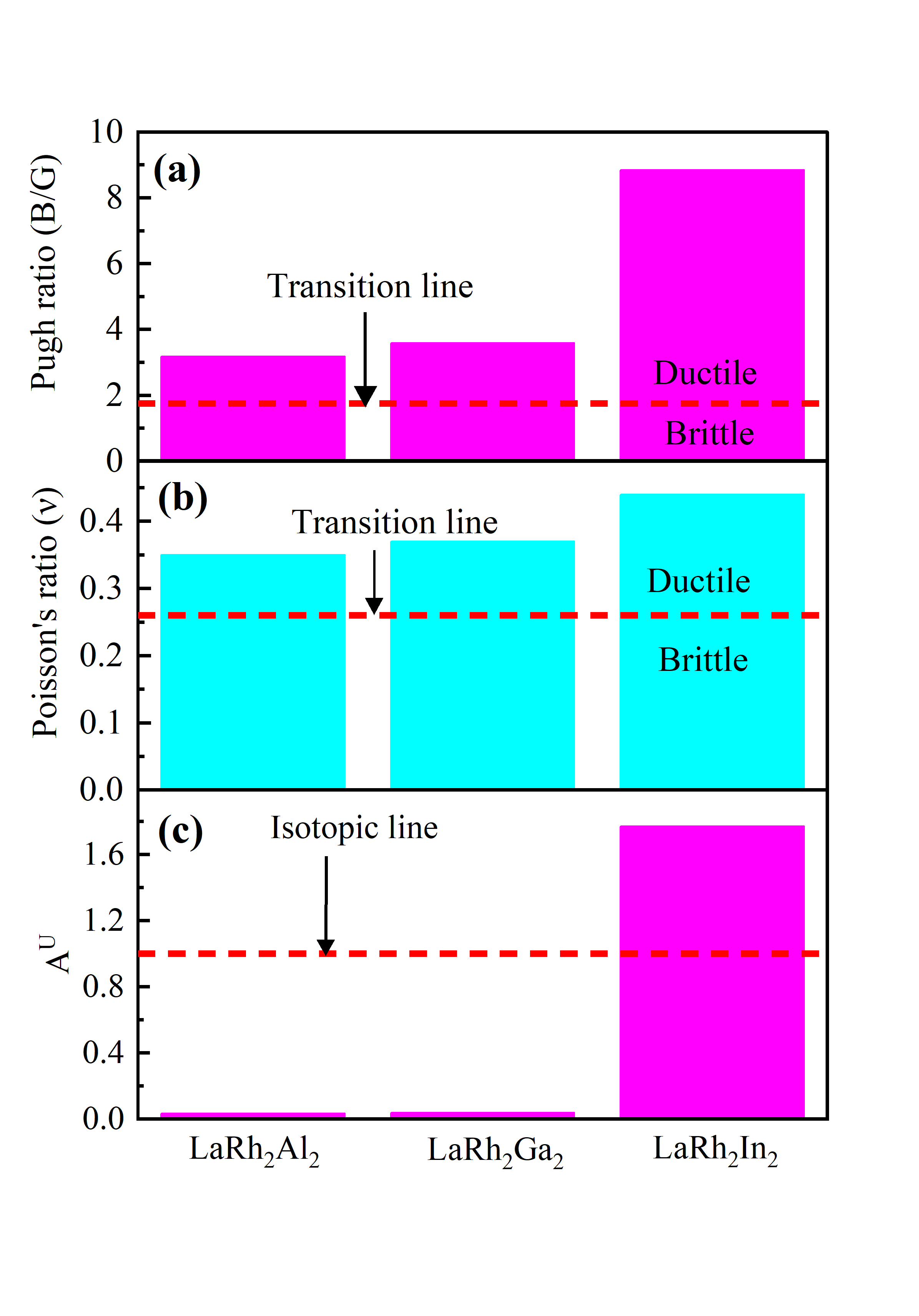}
		\caption{\label{mecha}(a) Pugh’s ratio ($B/G$), (b) Poisson ratio ($\nu$), and (c) Anisotropy factor of LaRh$_2$Al$_2$, LaRh$_2$Ga$_2$, and LaRh$_2$In$_2$ superconductors.
		}
	\end{figure}
	Another important parameter, named the anisotropy factor ($A$), shows a value of 1.0 for denoting isotropy, whereas values deviating from 1.0 indicate anisotropy, as shown in Fig.~2(c), which provides a graphical representation as well. The elastic anisotropy of the material was analyzed by estimating the direction-dependent variations in Young's modulus, compressibility, shear modulus, and Poisson's ratio using the ELATE code~\cite{c31}. Fig.~3 presents 3D contour plots of these elastic properties, where spherical shapes represent isotropic behavior and deviations indicate anisotropy. The plots reveal slight deviations from a spherical shape, suggesting a mild degree of anisotropy in the materials.
	Moreover, when investigating the various bonding natures for specific crystallographic directions, the anisotropy factor is a crucial consideration. Understanding and improving the mechanical behavior of crystalline materials requires an understanding of the anisotropy factor and elastic anisotropy.
	\begin{equation}
		A^U = \frac{5G_V}{G_R} + \frac{B_V}{B_R} - 6
		\tag{10}
	\end{equation}
	
	A values of 0.033, 0.04, and 1.76 for LaRh$_2$Al$_2$, LaRh$_2$Ga$_2$, and LaRh$_2$In$_2$, respectively, confirm the anisotropic nature of all the studied materials.
	\begin{figure}[ht]
		\centering
		\includegraphics[width=0.6\textwidth]{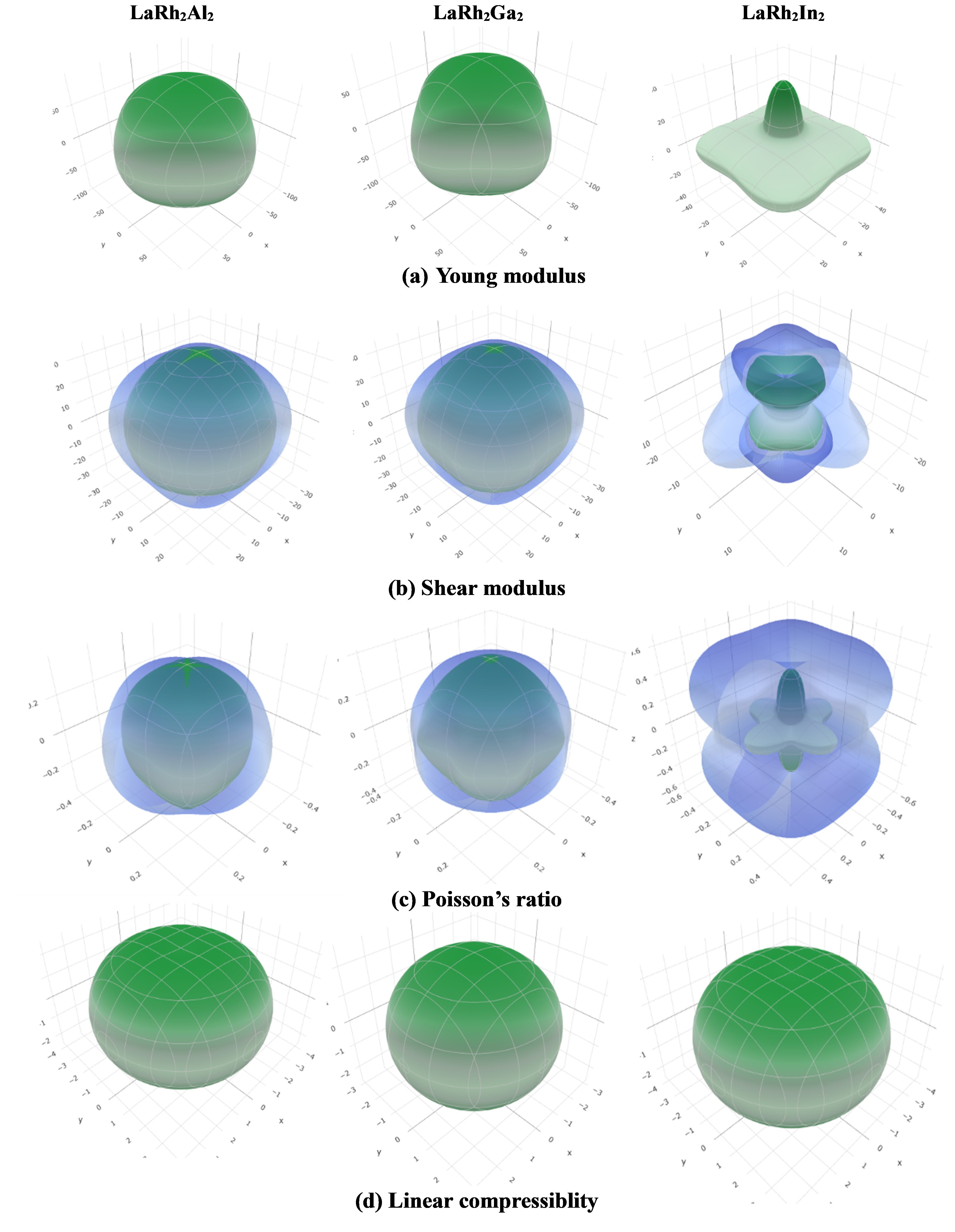}
		\caption{\label{ddha}
			3D directional dependences of (a) Young's modulus, (b) shear modulus, and (c) Poisson’s ratio of LaRh$_2$Al$_2$, LaRh$_2$Ga$_2$, and LaRh$_2$In$_2$.
		}
	\end{figure}
	
	Notably, the projections of Young’s modulus (Y), compressibility (K), shear modulus (G), and Poisson’s ratio in the ab-plane appear nearly circular. This indicates that elastic anisotropy within the basal plane is minimal. Table 5 shows the maximum and minimum values of Y, K, and G, as well as their ratios, to provide quantitative measurements of elastic anisotropy. The results show the presence of anisotropy, while it is still quite modest. However, this study is critical for understanding the material's mechanical behavior in different directions.

	\begin{table}[htbp]
		\centering
		\caption{The minimum and maximum limits of Young's modulus (Y), Shear modulus (G), Poisson's ratio ($\nu$), and Linear compressibility ($\beta$) of LaRh$_2$X$_2$ (X = Al, Ga, In).}
		\label{tab:limits_moduli}
		\begin{tabular}{llccc}
			\toprule
			& & \textbf{LaRh$_2$Al$_2$} & \textbf{LaRh$_2$Ga$_2$} & \textbf{LaRh$_2$In$_2$} \\
			\midrule
			\multirow{2}{*}{Young's modulus (GPa)} 
			& $E_\mathrm{min}$ & 88.02  & 81.615 & 19.267 \\
			& $E_\mathrm{max}$ & 101.81 & 97.496 & 53.329 \\
			\midrule
			\multirow{2}{*}{Shear modulus (GPa)} 
			& $G_\mathrm{min}$ & 32.779 & 29.158 & 5.5284 \\
			& $G_\mathrm{max}$ & 40.504 & 38.004 & 21.738 \\
			\midrule
			\multirow{2}{*}{Poisson's ratio} 
			& $\nu_\mathrm{min}$ & 0.25676 & 0.2827  & 0.13584 \\
			& $\nu_\mathrm{max}$ & 0.43129 & 0.42376 & 0.79344 \\
			\midrule
			\multirow{2}{*}{Linear compressibility (TPa$^{-1}$)} 
			& $\beta_\mathrm{min}$ & 2.3884 & 2.5256 & 2.3736 \\
			& $\beta_\mathrm{max}$ & 3.0641 & 3.0108 & 4.0602 \\
			\bottomrule
		\end{tabular}
	\end{table}
	
	\subsubsection{Debye Temperature and Relevant Properties}
	At a particular temperature, the greatest frequency mode of vibration within a solid is represented by the Debye temperature. It is crucial for determining a solid's melting point, specific heat, thermal expansion, among other physical characteristics. Furthermore, the Debye temperature aids in explaining how materials behave in both high and low temperature ranges. Understanding the thermal dynamics and vibrational characteristics of solids requires an understanding of this parameter. The Debye temperature influences important material properties and provides information on the thermophysical behavior of solids. Moreover, it serves as a boundary between the classical and quantum behavior of phonons. When the temperature ($T$) of a solid exceeds the Debye temperature ($\theta_\mathrm{D}$), each vibrational mode is expected to have an energy of $k_\mathrm{B}T$. However, at temperatures below $k_\mathrm{B}T < \theta_\mathrm{D}$, high-frequency vibrational modes remain inactive. In such low-temperature conditions, the vibrational energy primarily comes from acoustic modes. This distinction highlights the shift in phonon behavior from quantum effects at low temperatures to classical behavior at higher temperatures. This approach provides a way to determine the Debye temperature using the average sound velocity, $v_m$. It relies on a specific equation that links the Debye temperature to $v_m$. This calculation method serves as a practical tool for studying a material’s diverse properties using the following equations.
	\begin{equation}
		\theta_D = \frac{h}{K_B} \left[ \frac{3n}{4\pi} \left( \frac{N_A \rho}{M} \right) \right]^{1/3} v_m
		\tag{11}
	\end{equation}
	
	\begin{equation}
		v_m = \left[ \frac{1}{3} \left( \frac{2}{v_t^3} + \frac{1}{v_l^3} \right) \right]^{-1/3}
		\tag{12}
	\end{equation}
	
	\begin{equation}
		v_l = \left( \frac{B + \frac{4}{3}G}{\rho} \right)^{1/2}
		\tag{13}
	\end{equation}
	
	\begin{equation}
		v_t = \left( \frac{G}{\rho} \right)^{1/2}
		\tag{14}
	\end{equation}
	Where, $K_B$, $h$, $N_A$, $M$, $\rho$, and $n$ indicate the Boltzmann constant, the Planck constant, Avogadro’s number, the molecular mass, the density, and the number of atoms in the unit cell, respectively. A lower Debye temperature is typically associated with decreased phonon thermal conductivity.
	
	Among the studied compounds, the LaRh$_2$In$_2$ superconductor has a much lower Debye temperature than the other two isostructural materials. This characteristic makes LaRh$_2$In$_2$ a viable candidate for applications requiring thermal barrier coatings due to its reduced thermal conductivity. Given these findings, LaRh$_2$X$_2$ shows potential to use as a thermal barrier coating material. This result is supported by the relative low value of the Young's modulus of the LaRh$_2$X$_2$ as well. Its properties align with the requirements for materials designed to minimize heat transfer.
	\begin{table}[htbp]
		\centering
		\caption{The calculated density ($\rho$), Debye temperature ($\theta_D$), and melting temperature ($T_m$) along with the longitudinal ($v_l$), transverse ($v_t$), and average ($v_m$) sound velocities determined using the various expressions.}
		\begin{tabular}{lccccccc}
			\hline
			\textbf{Compounds} & $\theta_D$ (K) & $\rho$ (g/cm$^3$) & $v_t$ (km/sec) & $v_l$ (km/sec) & $v_m$ (km/sec) & $T_m$ (K) & \textbf{Ref.} \\
			\hline
			LaRh$_2$Al$_2$ & 282.11 & 6.79 & 2.27 & 4.82 & 2.55 & 1070.47 & This \\
			LaRh$_2$Ga$_2$ & 248.15 & 8.24 & 1.99 & 4.41 & 2.244 & 1090.57 & This \\
			LaRh$_2$In$_2$ & 133.40 & 9.78 & 1.05 & 3.35 & 1.27 & 872.97 & This \\
			\hline
		\end{tabular}
	\end{table}
	Melting point of solids is the temperature at which it transitions from a solid to a liquid under normal atmospheric pressure. The elastic constants $C_{ij}$ can be used to calculate the melting temperature of a tetragonal crystals. The relation is represented by a particular equation designed for these kinds of calculations.
	
	\begin{equation}
		T_m = 354 + 4.5 \frac{2C_{11} + C_{33}}{3}
		\tag{15}
	\end{equation}
	\sloppy

	Here, the melting temperature $T_m$ is measured in K, while $C_{11}$ and $C_{33}$ are expressed in GPa. The calculated melting temperatures for LaRh$_2$Al$_2$, LaRh$_2$Ga$_2$, and LaRh$_2$In$_2$ are 282.11~K, 248.15~K, and 133.40~K, respectively, as shown in Table~5. This analysis shows that LaRh$_2$Al$_2$ exhibits the highest melting temperature among the compounds studied, surpassing that of the two superconductors.
	
	\subsubsection{Phonon dispersion and Thermodynamic Properties}
	Phonon dispersion exerts a significant role in the physical characteristics and dynamic behavior of materials, including phase transition, elastic properties, thermal conductivity, electron-phonon interactions, vibrational modes, Raman and IR active modes, and thermal expansion~\cite{c311}. In the first Brillouin zone of the tetragonal structure of LaRh$_2$Al$_2$, LaRh$_2$Ga$_2$ and LaRh$_2$In$_2$ superconductors, the phonon dispersion band and density of states along high symmetry points have been investigated \cite{c32}. The vibrationally dynamic stability of these compounds is supported by a positive dispersion curve. Fig.~4(a-c) indicate that the phonon dispersion curves for LaRh$_2$Al$_2$ and LaRh$_2$Ga$_2$ superconductors have no negative values, suggesting that they are both dynamically stable, whereas LaRh$_2$In$_2$ has a tiny number of negative values, showing that it has slight dynamical instability in the lattice \cite{c33}. The phonon partial density of states (PDOS) study shows that the vibrations of the Rh--Al, Rh--Ga, and Rh--In bonds in the three corresponding compounds are the main source of the low-frequency range (1--3~THz), with the heavier Rh atoms having the most influence on the vibrational modes. The Rh--La bonds are mostly responsible for the vibrations in the mid-frequency region (4--5~THz). Interestingly, the LaRh$_2$In$_2$ compound's phonon spectrum contains imaginary (negative) frequencies, indicating dynamic instability that could point to a structural phase shift. The Rh atom is mainly responsible for this instability, underscoring its important function in the observed negative-frequency modes. A significant energy shift at the Z-$\Gamma$ point suggests the longitudinal acoustic mode. Optical modes and acoustic modes overlap for all of the compounds. Characteristics data including Kohn anomalies, significant dips in scatter from Z to $\Gamma$ point, may also indicate strong electron-phonon interaction, which is essential to superconductive characteristics. The $\Gamma$-point congruence of optical phonons may help with spectroscopic assessment by forecasting both IR-active and Raman-active modes.
	\begin{figure}[ht]
		\centering
		\includegraphics[width=1.0\textwidth]{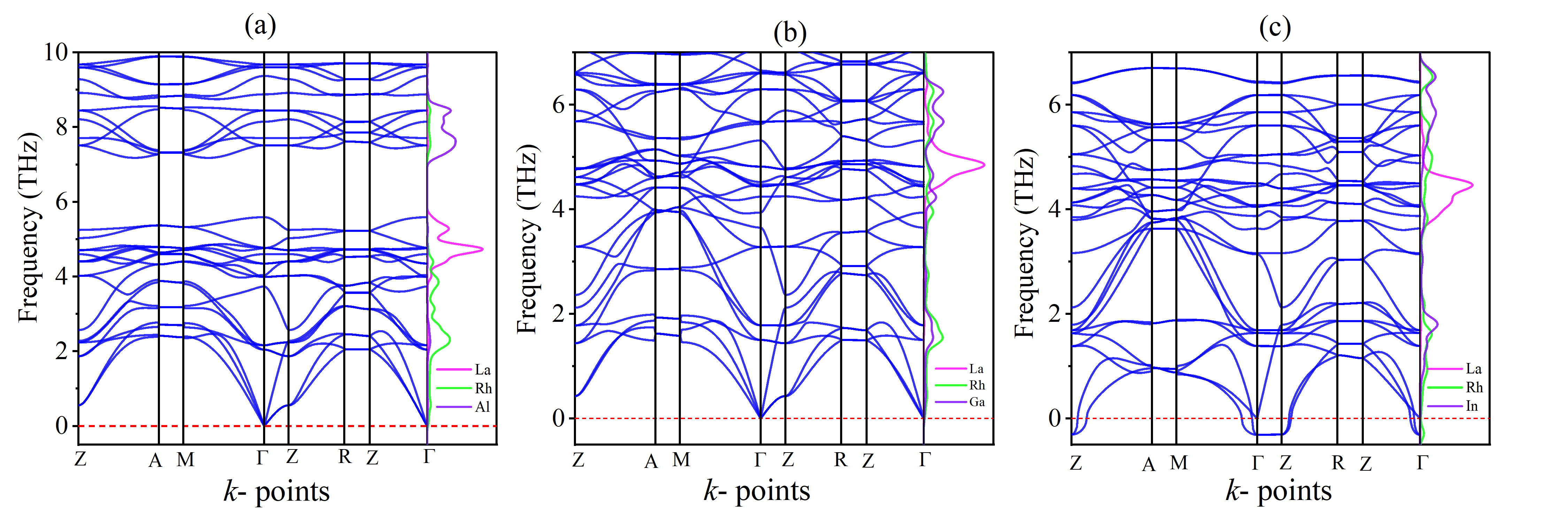}
		\caption{\label{dd1}
			Calculated phonon spectra for (a) LaRh$_2$Al$_2$, (b) LaRh$_2$Ga$_2$, and (c) LaRh$_2$In$_2$. For each compound, the phonon dispersion relation is shown on the left, and the corresponding phonon density of states (DOS) is displayed on the right.
		}
	\end{figure}
	\\The thermodynamic properties of LaRh$_2$X$_2$ (X = Al, Ga, In) compounds are analyzed as a function of temperature shown in Fig.~5. The plots show variations in enthalpy, free energy, and entropy for each compound over a temperature range from 0 to 1000~K. Enthalpy increases gradually with temperature for all superconductors, indicating heat absorption as the system's energy changes \cite{c35}.
	Free energy decreases with temperature, reflecting the increasing spontaneity of the compounds at higher temperatures. Whereas entropy robustly increases with temperature, showing that the disorder within the system rises as temperature increases. The zero-point energy values for LaRh$_2$Al$_2$, LaRh$_2$Ga$_2$, and LaRh$_2$In$_2$ are 0.3527~eV, 0.2775~eV, and 0.2426~eV, respectively, indicating their intrinsic energy levels. The heat capacity ($C_p$) graph shows similar behavior for all three superconductors, reaching saturation around 60~cal/(cell$\cdot$K) after 400~K. The heat capacity curve demonstrates a rapid increase at low temperatures, which gradually flattens as temperature rises. The thermodynamic stability of these compounds can be inferred from the trends in their free energy and entropy values.
	\begin{figure}[ht]
		\centering
		\includegraphics[width=0.6\textwidth]{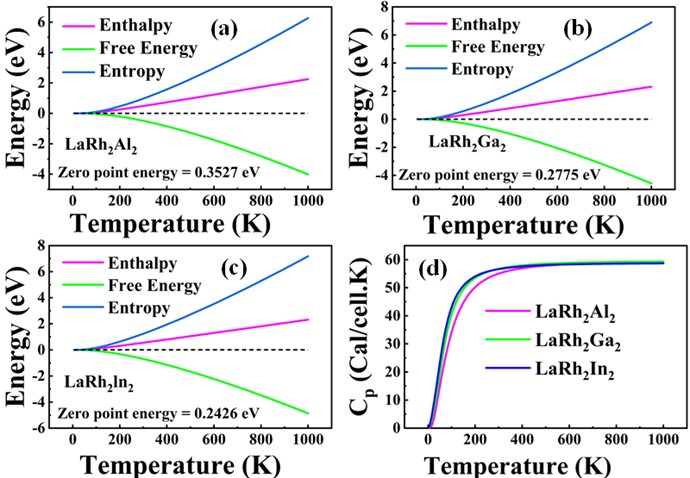}
		\caption{\label{dd2}
			Calculated thermodynamic properties for (a) LaRh$_2$Al$_2$, (b) LaRh$_2$Ga$_2$, and (c) LaRh$_2$In$_2$
		}
	\end{figure}
	\subsubsection{Theoretical Vicker’s Hardness (H$_v$)}
	Hardness measures a material’s ability to resist localized plastic deformation caused by mechanical indentation or abrasion. It provides an estimate of how well a material withstands such external forces. Both external and intrinsic factors influence the hardness of a material. Features such as cohesive energy, binding strength, and crystal structure are examples of intrinsic qualities. Conversely, extrinsic features include things like morphology and stress fields.
	
	Temperature and other environmental factors, as well as the measuring methods used, might affect experimental hardness values. Similarly, the particular models and computation techniques employed have an impact on theoretical hardness values. These variations highlight the complexity in precisely defining and measuring hardness~\cite{c36}. Understanding both intrinsic and extrinsic contributions is essential for a comprehensive assessment of material hardness. The Vicker’s hardness $H_\nu$ of a compound is determined using Mulliken bond population data. This calculation is performed through specific mathematical relations~\cite{c37}.
	\begin{align}
		H_v^{\mu} &= 740 (P^{\mu} - P^{\mu''}) (V_b^{\mu})^{-\frac{5}{3}} \tag{17} \\[1em]
		H_v &= \left[ \prod_{\mu} (H_v^{\mu})^{n^{\mu}} \right]^{1/\sum n^{\mu}} \tag{18} \\[1em]
		V_b^{\mu} &= (d^{\mu})^3 \Big/ \sum_{\nu} \left[ (d^{\mu})^3 N_b^{\nu} \right] \tag{19} \\[1em]
		P^{\mu''} &= \frac{n_{\text{free}}}{V} \tag{20}
	\end{align}
	
	where, $P^{\mu}$ is called the Mulliken population of the $\mu$-type bond, $P^{\mu''}$ denotes the metallic population of the $\mu$-type bond, $n_{\text{free}}$ is the number of free electrons, $V_b^{\mu}$ is the volume of a bond of type $\mu$, $V$ denotes the cell volume, $d^{\mu}$ is the bond length of type $\mu$ and $N_b^{\nu}$ is called the bond number of type $\nu$ per unit volume. Diamond is recognized as the hardest material, with a Vickers hardness ranging from 70 to 150~GPa. According to \textbf{Table~6}, the calculated Vickers hardness values for the superconductors are LaRh$_2$Al$_2$, LaRh$_2$Ga$_2$ and LaRh$_2$In$_2$ are 6.23~GPa, 2.28~GPa, and 3.75~GPa, respectively.
	\begin{table}[htbp]
		\centering
		\caption{Calculated Mulliken bond overlap population of $\mu$-type bond $P^{\mu}$, $n^{\mu}$ is the number of bonds, bond length $d^{\mu}$, metallic population $P^{\mu''}$, bond volume $V_b^{\mu}$ (Å$^3$) and Vickers hardness of $\mu$-type bond $H_b^{\mu}$ (GPa) and total hardness $H_v$ (GPa) of LaRh$_2$Al$_2$, LaRh$_2$Ga$_2$, and LaRh$_2$In$_2$ superconductors.}
		\begin{tabular}{llcccccccc}
			\toprule
			\textbf{Compound} & \textbf{Bond} & $n^{\mu}$ & $d^{\mu}$ (\AA) & $P^{\mu}$ & $P^{\mu''}$ & $V_b^{\mu}$ (Å$^3$) & $H_b^{\mu}$ (GPa) & $H_v$ (GPa) & \textbf{Ref.} \\
			\midrule
			\multirow{3}{*}{LaRh$_2$Al$_2$} 
			& Rh--Al & 4 & 2.49 & 1.27 &      & 18.39  & 7.23 & \multirow{3}{*}{This} \\
			& Rh--Al & 4 & 2.54 & 1.31 & 0.019 & 19.52  & 6.75 &                      \\
			& Rh--Al & 2 & 2.59 & 0.85 &      & 20.70  & 3.94 &                      \\
			\midrule
			\multirow{3}{*}{LaRh$_2$Ga$_2$} 
			& Rh--Ga & 4 & 2.52 & 0.43 &      & 19.18  & 2.22 & \multirow{3}{*}{This} \\
			& Rh--Ga & 4 & 2.53 & 0.58 & 0.0181 & 19.65 & 2.91 &                      \\
			& Rh--Ga & 2 & 2.54 & 0.31 &      & 19.74  & 1.50 &                      \\
			\midrule
			\multirow{3}{*}{LaRh$_2$In$_2$} 
			& Rh--In & 4 & 2.71 & 1.04 &      & 23.65  & 3.89 & \multirow{3}{*}{This} \\
			& Rh--In & 4 & 2.67 & 1.08 & 0.0142 & 23.003 & 4.24 &                      \\
			& Rh--In & 2 & 2.63 & 0.64 &      & 21.86  & 2.71 &                      \\
			\bottomrule
		\end{tabular}
	\end{table}
	
	These values indicate that the studied materials are significantly softer than diamond. As a result, their resistance to deformation is relatively low. This comparison underscores the soft nature of these superconducting compounds and is comparable to previous reports.
	\subsection{Electronic properties}
	\subsubsection{Band Structure}
	The electronic band structure is a fundamental idea in solid-state physics that characterizes various materials' electrical conductivity, thermal conductivity, electronic heat capacity, hall effect, magnetic properties, and optoelectronic properties. Optical and charge transport properties are almost completely defined by the band structure. For LaRh$_2$Al$_2$, LaRh$_2$Ga$_2$, and LaRh$_2$In$_2$, we have computed the electronic band structures using DFT calculations, following structural optimization with a k-points grid of $9\times9\times4$ in specific directions within the Brillouin zone. The band structures of the studied compounds are computed along the high point symmetry in the first Brillouin zone in the $\Gamma$-M-K-$\Gamma$-A-L-H-A path as depicted in \textbf{Figures 6(a-c)}. All three compounds have no bandgap and significant band overlapping at the Fermi level ($E_F$), indicating their metallic operations.
	The overlap between the valence and conduction bands allows unrestricted electron mobility, which contributes to higher electrical conductivity. The band structures exhibit considerable hybridization of La, Rh, and In atomic orbitals, resulting in bonding and antibonding states. As a result of this hybridization, bonding and antibonding states are forming and shaping the electronic structure of these superconducting materials.
	\begin{figure}[ht]
		\centering
		\includegraphics[width=0.98\textwidth]{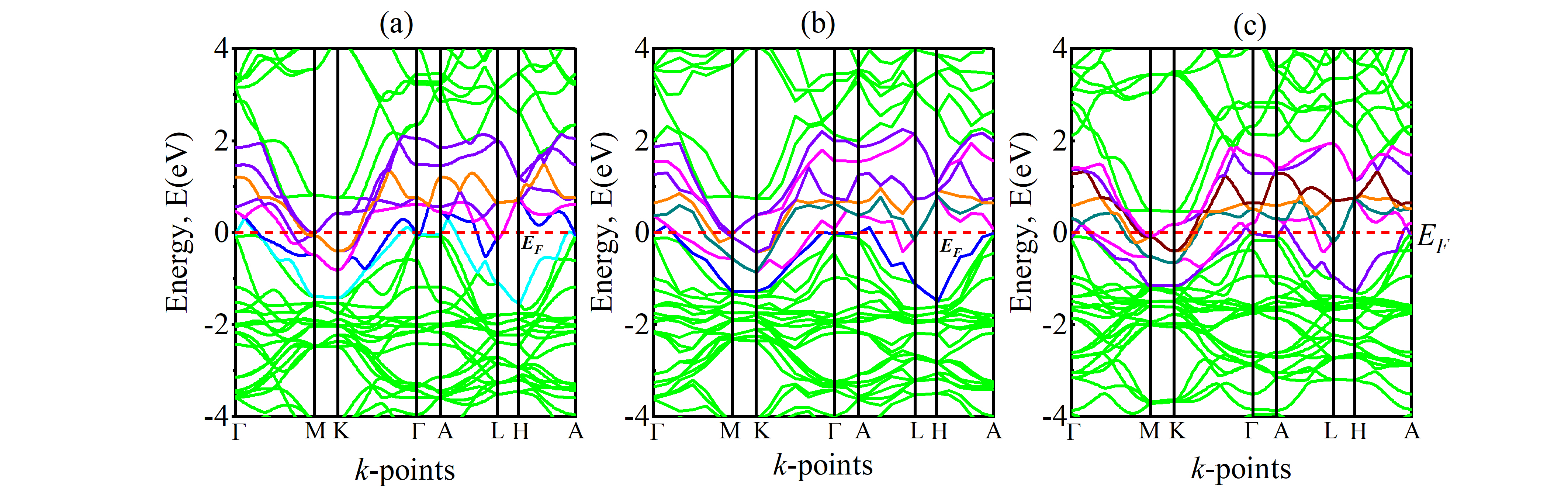}
		\caption{\label{dd3}
			The energy band structures of (a) LaRh$_2$Al$_2$, (b) LaRh$_2$Ga$_2$, and (c) LaRh$_2$In$_2$ calculated by GGA-PBE approximation.
		}
	\end{figure}
	
	\subsubsection{Density of states (DOS) }
	The electronic energy density of states (DOS) is a pivotal parameter in solid-state physics, delineating the distribution of accessible electronic states per unit energy across various energy levels of band structure. Each atomic constituent contributes distinctively to both bonding and antibonding states, profoundly shaping the material’s electronic characteristics. The calculated the total and partial density of states (TDOS and PDOS) for LaRh$_2$X$_2$ (X = Al, Ga, In) incorporating detailed orbital-specific contributions, as depicted in Fig.~7 over an energy range spanning from $-8$~eV to $+8$~eV, employing the Gaussian smearing method.
	
	\begin{table}[ht]
		\centering
		\caption{Total DOS and Partial DOS value at different states of superconductors.}
		\begin{tabular}{lccc}
			\toprule
			\textbf{TDOS/PDOS} & \textbf{LaRh$_2$Al$_2$} & \textbf{LaRh$_2$Ga$_2$} & \textbf{LaRh$_2$In$_2$} \\
			\midrule
			TDOS & 5.75 & 5.52 & 6.2 \\
			\midrule
			\multicolumn{4}{l}{\textbf{PDOS}} \\
			La-5d   & 1.41 & 1.30 & 1.49 \\
			Rh-4p   & 1.49 & 1.60 & 1.86 \\
			Rh-4d   & 1.44 & 1.13 & 1.23 \\
			Al-3p   & 1.33 &      &      \\
			Ga-4p   &      & 1.31 &      \\
			In-5p   &      &      & 1.32 \\
			\bottomrule
		\end{tabular}
	\end{table}
	
	The Fermi level ($E_F$), represented by a vertical dashed line, serves as a fundamental reference point in energy space. According to \textbf{Fig.~7} the presence of non-zero TDOS values at $E_F$ clearly confirms the metallic nature of these compounds. Specifically, the calculated TDOS values at $E_F$ for LaRh$_2$Al$_2$, LaRh$_2$Ga$_2$, and LaRh$_2$In$_2$ are 5.75, 5.52, and 6.22 states/eV, respectively, emphasizing their intrinsic electronic disparities. Furthermore, an in-depth PDOS analysis for La, Rh, Al, Ga, and In atoms elucidates their discrete contributions to the TDOS and their role in modulating chemical bonding interactions. The predominant contributions to the TDOS in proximity to $E_F$ originate from La-5$d$ states, Rh-4$p$ and Rh-4$d$ states, alongside the 3$p$/4$p$/5$p$ states of Al, Ga, and In, which vary across LaRh$_2$Al$_2$, LaRh$_2$Ga$_2$, and LaRh$_2$In$_2$, respectively. The valence band, extending from $-4.72$~eV to $-0.8$~eV, is largely governed by Rh-4$d$ states, underscoring their fundamental role in dictating the chemical and mechanical resilience of LaRh$_2$X$_2$ compounds. Additionally, the valence band near $E_F$ ($-0.8$~eV to 0~eV) exhibits a pronounced hybridization effect between Rh-4$d$ states and the 3$p$/4$p$/5$p$ states of Al, Ga, and In, depending on the specific LaRh$_2$X$_2$ compound. Notably, the primary peak positions and the relative alignment of $E_F$ exhibit pronounced sensitivity to structural perturbations such as lattice distortions, elemental doping, and applied stress. The electronic and structural stability of these compounds is intrinsically contingent upon the precise positioning of $E_F$ concerning the surrounding TDOS. This comprehensive analysis reveals that LaRh$_2$In$_2$, with its highest states/eV at $E_F$, demonstrates superior electrical conductivity compared to LaRh$_2$Al$_2$ and LaRh$_2$Ga$_2$, establishing it as the most conductive among the studied compounds.
	\begin{figure}[ht]
		\centering
		\includegraphics[width=0.7\textwidth]{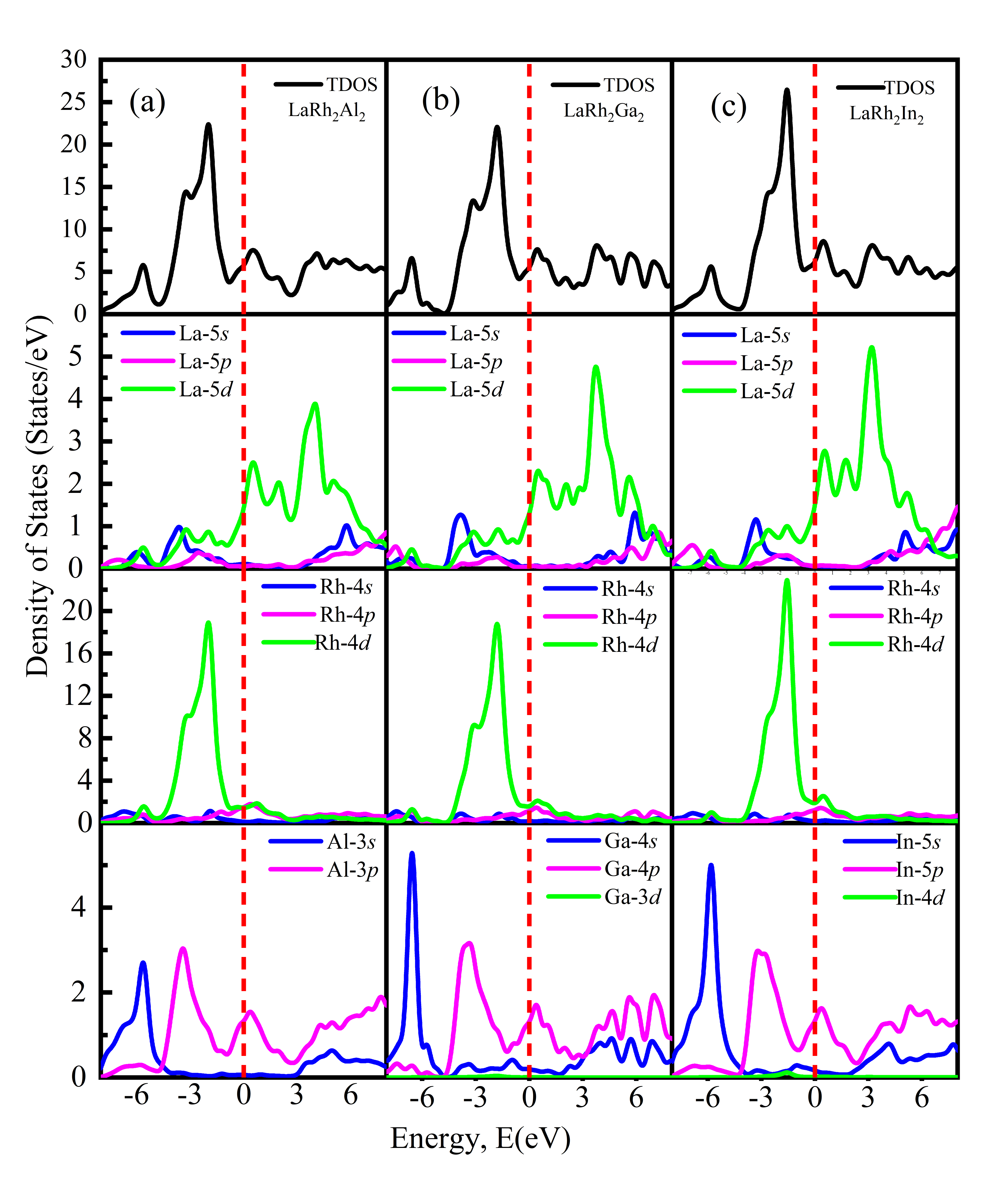}
		\caption{\label{dd4}
			The total and partial electronic density of states (DOS) for (a) LaRh$_2$Al$_2$, (b) LaRh$_2$Ga$_2$, and (c) LaRh$_2$In$_2$ as a function of energy.
		}
	\end{figure}
	\subsubsection{Density of states and Band structure (spin polarized calculation) }
	Spin-polarized calculations were also conducted to determine the spin-dependent density of states (DOS) and evaluate the magnetic contribution of orbital electrons in these materials. The total and partial densities of states for LaRh$_2$Al$_2$, LaRh$_2$Ga$_2$, and LaRh$_2$In$_2$ were computed, considering the spin-up and spin-down contributions. Figure~8 displays the distribution of electronic states as a function of energy, highlighting contributions from $s$, $p$, and $d$ orbitals. The Fermi level is set at 0~eV, indicated by the vertical red dashed line. The DOS plot for LaRh$_2$Al$_2$ reveals prominent peaks at the $E_F$, predominantly from Al-3$p$ orbital than others as shown in Fig.~8(a), with total DOS values of 2.60~states/eV for spin-up and $-2.60$~states/eV for spin-down, indicating a symmetric spin distribution. In LaRh$_2$Ga$_2$, the DOS near the Fermi level is slightly enhanced compared to LaRh$_2$Al$_2$, primarily due to Rh-4$d$ orbital contributions, with TDOS of 2.96~states/eV for spin-up and $-2.96$~states/eV for spin-down as depicted in Fig.~8(b). Whereas TDOS values are 3.49~states/eV for spin-up and $-3.49$~states/eV for spin-down with Rh-4$d$ dominant contribution, confirming symmetric spin contribution which might occur for the nonmagnetic or diamagnetic behavior of all compounds. 
	This expulsion of magnetic fields means that the material cannot sustain any internal magnetic moment, which is the behavior of superconductors~\cite{c41}.Interestingly, total and partial orbital spin contribution was much lower compared to non spin orbital contribution for all superconductors which indicates the spin electronic effects (SOC-spin orbit coupling) on these materials. 
	\begin{figure}[ht]
		\centering
		\includegraphics[width=0.8\textwidth]{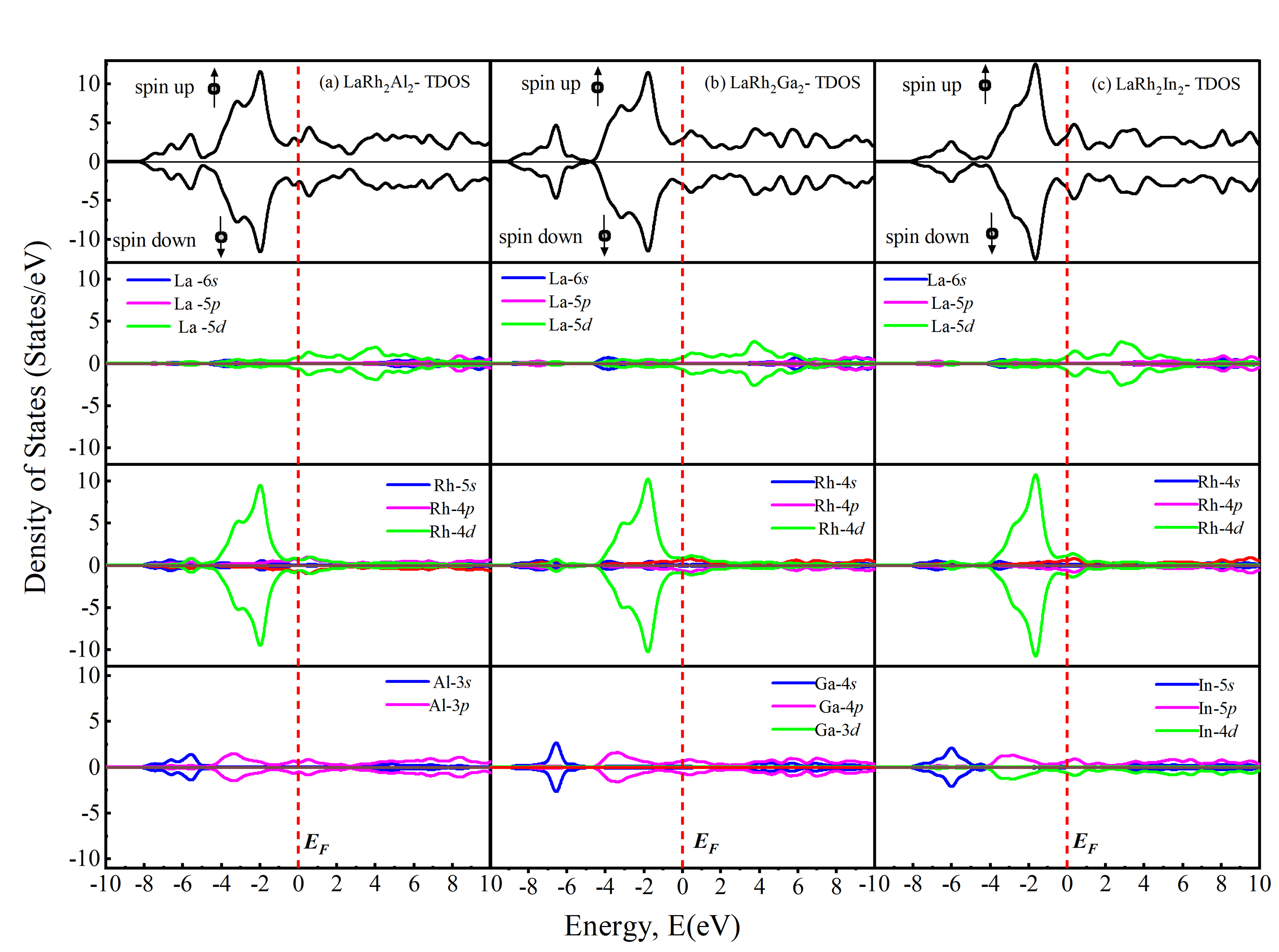}
		\caption{\label{dd5}
			Spin-resolved total density of states (TDOS) for (a) LaRh$_2$Al$_2$, (b) LaRh$_2$Ga$_2$, and (c) LaRh$_2$In$_2$ as a function of energy.
		}
	\end{figure}

	\begin{table}[ht]
		\centering
		\caption{Total density of states for LaRh$_2$X$_2$ (X = Al, Ga, In) obtained from spin-polarized calculations.}
		\begin{tabular}{llc}
			\toprule
			\textbf{Compound} & \textbf{Spin} & \textbf{Total DOS (states/eV)} \\
			\midrule
			LaRh$_2$Al$_2$ & Up   &  2.60 \\
			LaRh$_2$Al$_2$ & Down & -2.60 \\
			LaRh$_2$Ga$_2$ & Up   &  2.96 \\
			LaRh$_2$Ga$_2$ & Down & -2.96 \\
			LaRh$_2$In$_2$ & Up   &  3.49 \\
			LaRh$_2$In$_2$ & Down & -3.49 \\
			\bottomrule
		\end{tabular}
	\end{table}
	
	\subsubsection{Fermi Surface Analysis ) }
	To understand the behavior of electronic states, the study of the Fermi surface of metallic materials is essential. The Fermi surface is good to predict several properties of metals, semi-metals and doped semiconductors. Any material's conductivity can be understood by utilizing the size and shape of the Fermi surface. The calculated Fermi surface is illustrated in \textbf{Fig.~9} for (a) LaRh$_2$Al$_2$, (b) LaRh$_2$Ga$_2$, and (c) LaRh$_2$In$_2$, providing a 3D cross-sectional view of the Brillouin zone with different Fermi sheets rendered in gray and blue in visualizing their electrical properties. Pink arrows denote high-symmetry directions ($\Gamma$, X, Z, etc.), helping us understand the orientation of the Fermi pockets. All three compounds display complex, multi-sheet Fermi surfaces, suggesting metallic behavior with multiple bands crossing the $E_F$. The Fermi surface band graphic has a complex mixed character since the high-dispersive band and the quasiflat band both intersect the Fermi level at the same time. The multi-sheet Fermi surface consists of two dimensional electron-like sheets in the corner of the Brillouin zone. There is a sheet which is a hole-like Fermi surface at the $\Gamma$-point. The hole-like Fermi surface surrounding the $\Gamma$-point is connected to a hole pocket around the X-point.
	There are also four half-tube-shaped wings cutting along its own axis on the highly complex core electron-like sheet. Each of these four wings maintains its Z-R orientation as it becomes a shape. Due to its multiband nature and robust hybridization at X site elements, these compounds LaRh$_2$X$_2$ (X = Al, Ga, In) exhibit superconducting tendencies with improved transport properties. 
	However, the Fermi surface of LaRh$_2$Al$_2$ (Figure~9a) is more fragmented with smaller electron/hole pockets, which implies more localized or less dispersive bands that might carry lower carrier concentration or lower effective dimensionality. In contrast, LaRh$_2$Ga$_2$ shows slightly more connected Fermi surface than in LaRh$_2$Al$_2$, where larger central and corner pockets suggest increased band dispersion or hybridization, which indicate higher conductivity compared to LaRh$_2$Al$_2$ superconductor. The largest and most continuous Fermi surfaces with more spherical and extended central pocket is seen for LaRh$_2$In$_2$ (Figure~9c). 
	The phenomenon indicates the high Fermi velocity and more free-electron-like behavior with higher carrier density and superior metallic conductivity for the material. Therefore, the evolution from Al $\rightarrow$ Ga $\rightarrow$ In shows a trend toward larger, more delocalized Fermi surfaces in this series of superconductor.
	\begin{figure}[ht]
		\centering
		\includegraphics[width=0.9\textwidth]{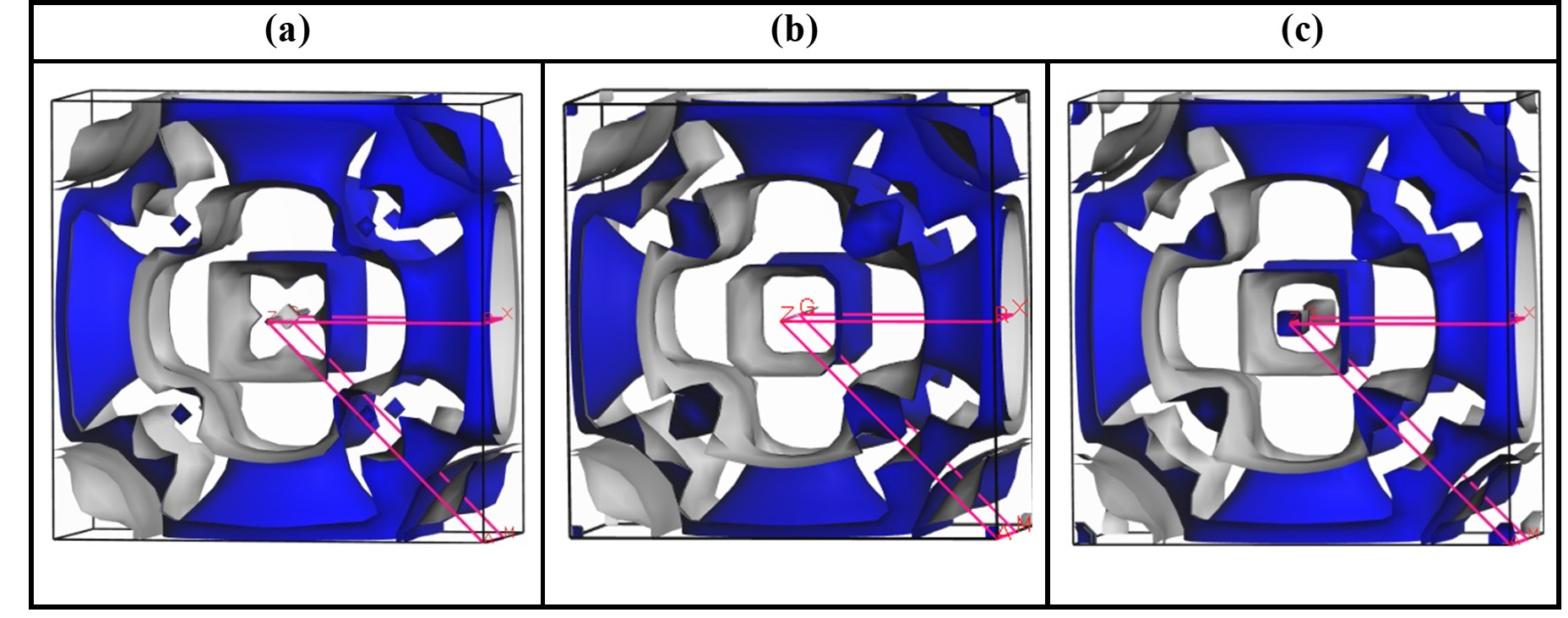}
		\caption{\label{dd65}The Fermi surface topologies of (a) LaRh$_2$Al$_2$, (b) LaRh$_2$Ga$_2$, and (c) LaRh$_2$In$_2$.
		}
	\end{figure}
	\subsubsection{Electron Charge Density Distribution}
	The type of bonding between the atoms is described by the electron density distribution. Investigating the charge density distribution is essential for identifying the bonding properties and explaining the charge transfer between the atoms. Fig.~10 shows the contour of the 2D electronic charge density distribution of LaRh$_2$X$_2$ (X = Al, Ga \& In) in the (100) plane. The color scale behind the figures' right side indicates the intensity of the electronic charge density, where red has low electron densities (0.0) while blue has high electron densities (maximum). The Mulliken bond overlap population study makes it clear that an ionic bonding between La and Rh is anticipated as a result of charge transfer from the La to Rh atom. The large electronegativity has a strong tendency to attract electrons toward itself~\cite{c42}. In the studied compounds, the electronegativity values are La~(1.1), Rh~(2.28), Al~(1.61), Ga~(1.81), and In~(1.78), respectively. The higher electronegativity of Rh~(2.28) is a strong accumulation of electron charge with Al/Ga/In, respectively.
	Additionally, weak charge buildup is seen at a lower electronegativity of La. The accumulation of electron charge between two atoms represents covalent bonds. In LaRh$_2$X$_2$ (X = Al, Ga, In), the strong covalent bonding is represented by Al--Rh, Ga--Rh, and Rh--In, respectively, which satisfy the Mulliken atomic and bond overlap population. Ionic bonding is indicated by the balance of positive and negative charges at the atomic location. This type of bonding characteristic of chemical can be explained by the oxidation states of the atoms. The compound's oxidation states---La$^{2+}$, Rh$^{1+}$, Al$^{3+}$, Ga$^{3+}$, and In$^{3+}$---indicate that it has no propensity for ionic bonding. Furthermore, because of the overlap of Rh 4$d$ states close to the Fermi level, metallic type bonding of the Rh--Rh link is predicted to exist. In LaRh$_2$X$_2$ (X = Al, Ga, In), the chemical bonding discussed above can be described as an anisotropic mix of ionic, metallic, and covalent interactions. Low charge density between Al and neighboring atoms (mostly red around Al) and very tiny localized charge at atomic sites La and Rh (small green rings) is visualized for LaRh$_2$Al$_2$. These scenarios indicate the minimal charge sharing between atoms with weak covalent bonding, but more ionic or metallic character. Increased charge accumulation around Ga compared to Al.
	\begin{figure}[ht]
		\centering
		\includegraphics[width=0.9\textwidth]{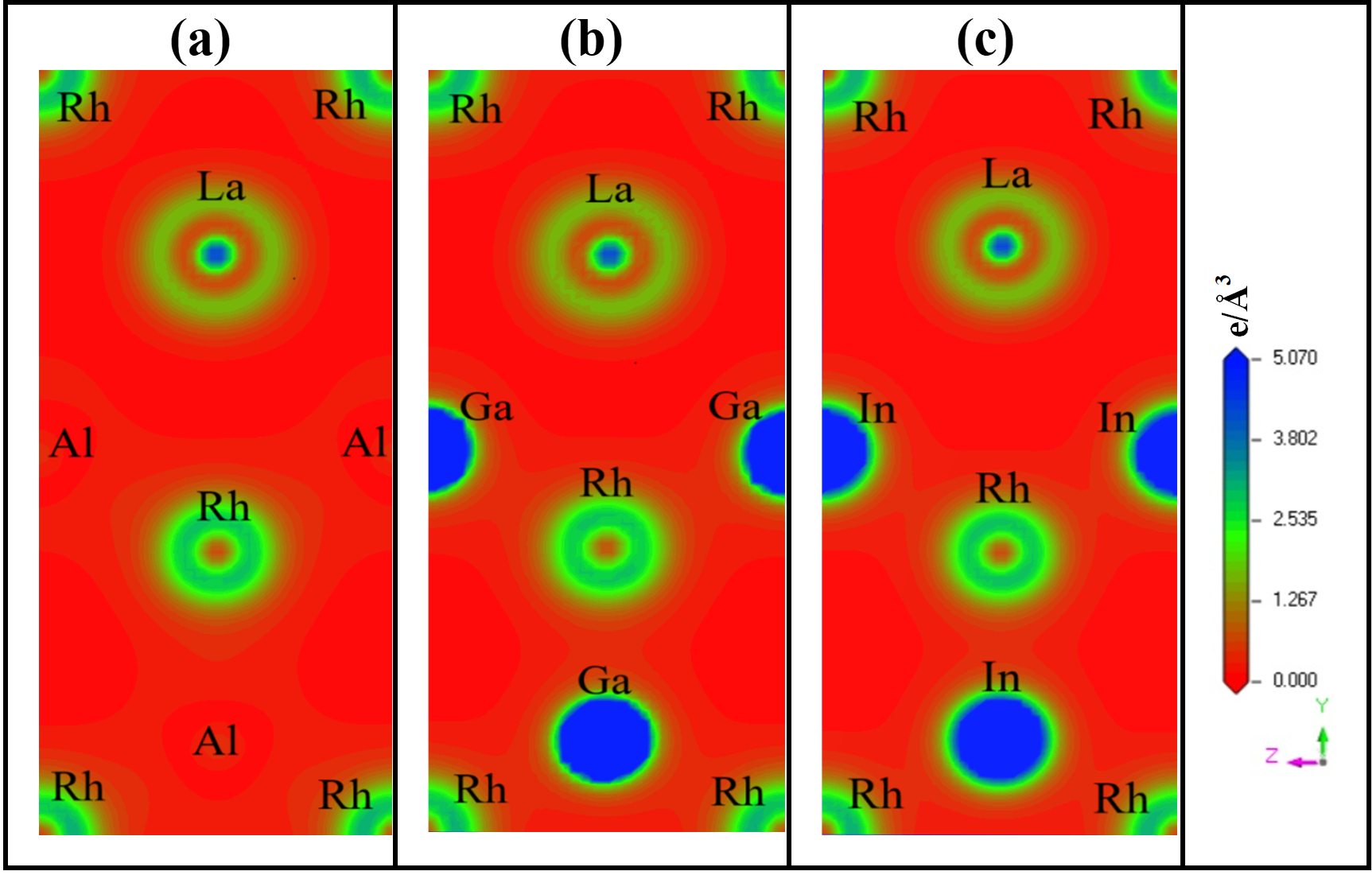}
		\caption{\label{dd6}Electronic charge density distribution maps of (a) LaRh$_2$Al$_2$, (b) LaRh$_2$Ga$_2$, and (c) LaRh$_2$In$_2$ in the (100) plane.
		}
	\end{figure}
	On the other hand, more blue-green spherical areas near Ga in LaRh$_2$Ga$_2$, indicating moderate charge delocalization and overlap with Rh which is an implication of stronger covalent interaction between Ga and neighboring Rh atoms. Highest charge density around In (deep blue), indicating significant electron localization. Bigger blue circular uniform regions between In and Rh atoms in LaRh$_2$Ga$_2$ suggest even greater charge overlap or stronger hybridization for more strong covalent character and potentially higher electronic polarizability. Therefore, the transition from Al to Ga to In represents the X-site element's increased p-orbital inclusion along with greater coupling with Rh-d orbitals, which presumably influences band dispersal and Fermi surface complication (as demonstrated in the previous picture). LaRh$_2$In$_2$ exhibits the largest charge concentration overlapping and a predominant metallic features amongst each of them. However, the higher hybrid and charge a delocalization match closely to the greater number of consistent and linked Fermi interfaces reported from Ga and In.
	\subsubsection{Mulliken atomic and bond overlap population }
	Mulliken population analysis is used to investigate the properties of chemical bonds and effective valence charge (EVC) that provide multiple functional explanations for the distribution of electrons among the different bond components. In order to analyze the population in the CASTEP code, Sanchez-Portal established a method that uses projection of the plane wave function that assigns charges in linear combination of atomic orbital (LCAO) basis sets \cite{c38}.
	The Mulliken charge assigned to a certain atomic species $a$ can be evaluated as:
	\setcounter{equation}{14}
	\begin{align}
		Q(\alpha) &= \sum_{k} w_k \sum_{\substack{\mu\, \text{on}\, \alpha}} \sum_{\nu} P_{\mu\nu}(k)\, S_{\mu\nu}(K) \tag{15} \\[2ex]
		P(\alpha\beta) &= \sum_{k} w_k \sum_{\substack{\mu\, \text{on}\, \alpha}} \sum_{\substack{\nu\, \text{on}\, \beta}} 2 P_{\mu\nu}(k)\, S_{\mu\nu}(K) \tag{16}
	\end{align}
	
	where $P_{\mu\nu}$ denotes the density matrix elements and $S_{\mu\nu}$ refers to the overlap matrix.
	The results for three investigated structures based on electronic charges are shown in Table~9. Effective valence charge (EVC) is the term used to describe the distinction between formal ionic charge and Mulliken charge. The EVC value close to zero indicates that ionic bond characteristics, and the value apart from zero indicates an increased level of covalency. The charge spilling parameter is represented by the quality of missing valence charge in a projection in the Mulliken charge population analysis. The lower values of the spilling parameter indicate a good representation of the electronic band. From the tables of charge spilling, LaRh$_2$Al$_2$ (0.35\%), LaRh$_2$Ga$_2$ (0.12\%), and LaRh$_2$In$_2$ (0.14\%), which indicates that LaRh$_2$Ga$_2$ superconductors have a good representation of electronic bonds.

	\begin{table}[ht]
		\centering
		\small 
		\caption{Charge spilling parameter (\%), atomic Mulliken charge, Hirshfeld charge, and effective valence charge of LaRh$_2$X$_2$ (X = Al, Ga, In).}
        \resizebox{\textwidth}{!}{%
        	\begin{tabular}{llccccccccc}
			\toprule
			\textbf{Compound} & \textbf{Charge spilling} & \textbf{Species} & \multicolumn{4}{c}{\textbf{Mulliken atomic populations}} & \textbf{Mulliken charge} & \textbf{Effective valence charge} & \textbf{Hirshfeld charge} & \textbf{Effective charge (EVC)} \\
			& \textbf{(\%)} &                 & s & p & d & Total & \textbf{charge} & \textbf{valence} & \textbf{charge} & \textbf{charge (EVC)} \\
			\midrule
			\multirow{3}{*}{LaRh$_2$Al$_2$} & 0.35 & La & 1.08 & 5.60 & 1.79 & 8.47 & 2.53 & 0.47 & 0.14 & 2.86 \\
			&      & Rh & 1.06 & 1.00 & 8.52 & 10.58 & -1.58 & 1.42 & -0.08 & 2.92 \\
			&      & Al & 1.13 & 1.77 & 0.00 & 2.90 & -0.10 & 2.90 & -0.05 & 2.95 \\
			\midrule
			\multirow{3}{*}{LaRh$_2$Ga$_2$} & 0.12 & La & 2.29 & 6.01 & 1.73 & 10.04 & 0.96 & 2.04 & 0.12 & 2.88 \\
			&      & Rh & 0.98 & 0.94 & 8.43 & 10.35 & -1.35 & 1.65 & -0.09 & 2.91 \\
			&      & Ga & 0.20 & 2.00 & 9.99 & 12.20 & 0.80 & 2.20 & 0.12 & 2.88 \\
			\midrule
			\multirow{3}{*}{LaRh$_2$In$_2$} & 0.14 & La & 1.24 & 6.30 & 1.67 & 9.20 & 1.80 & 1.20 & 0.14 & 2.86 \\
			&      & Rh & 0.90 & 0.77 & 8.50 & 10.17 & -1.17 & 1.83 & -0.15 & 2.85 \\
			&      & In & 1.08 & 1.83 & 9.97 & 12.88 & 0.12 & 2.88 & 0.14 & 2.86 \\
			\bottomrule
		\end{tabular}
        }
	
	\end{table}
	
	The Mulliken charge for La/Rh/Ga is 0.96/-1.35/0.80 in this case, meaning that the electronic charge is transferred to Rh atoms by the La and Ga atoms. We analyzed the bond overlap population ($P^{\mu}$) in order to comprehend the bonding properties of three superconductors (Table~10). The bond overlap population's empty value represents a completely ionic bonding, but the variation from zero indicates greater amounts of covalent bonds. It is observed from Table~10 that Al--Rh, Ga--Rh, and Rh--In exhibit robust covalent bonds and meet the Mulliken atomic and bond overlap population. Table~10 shows that the In--Rh link in LaRh$_2$In$_2$ appears more covalent compared to other superconductors. The $P^{\mu}$ figure is positive for the identical bond, but minus for X--La (X = Al, Ga, and In) and X--X bonds. Both the positive and negative values of $P^{\mu}$ represent the direct and indirect bonds between the elements engaged. At LaRh$_2$X$_2$ (X = Al, Ga, In), the bonds Al--Rh and X--La exhibit negative bond overlap populations, suggesting antibonding nature. Whereas in LaRh$_2$Ga$_2$, have better positive bond overlap populations, showing dominant covalent bonding. Ga--Ga and La--Ga bonds in the second coordinating layer cause antibonding and covalency in the material.
	
	In--Rh bonds in LaRh$_2$In$_2$ exhibit a much higher favorable bond overlap population than that of others. The presence of superconductive properties in the investigated superconductors might be due to varying bonding modes~\cite{c39}. The compound's partial ionic bonding between its various atoms is shown by the negative charge, which originates from the Ga-4p and La-5d states. A similar behavior is observed in LaRh$_2$Al$_2$ and LaRh$_2$In$_2$ structures. Therefore, it is evident from the discussion that the structures contain covalent bonds, which are already satisfied by the charge density distribution.\\
	\begin{tabular}{lccc lccc lccc}
    
		\toprule
		\multicolumn{4}{c}{\textbf{LaRh$_2$Al$_2$}} & \multicolumn{4}{c}{\textbf{LaRh$_2$Ga$_2$}} & \multicolumn{4}{c}{\textbf{LaRh$_2$In$_2$}} \\
		\cmidrule(lr){1-4} \cmidrule(lr){5-8} \cmidrule(lr){9-12}
		Bond & $n^{\mu}$ & $d^{\mu}$ (\AA) & $p^{\mu}$ &
		Bond & $n^{\mu}$ & $d^{\mu}$ (\AA) & $p^{\mu}$ &
		Bond & $n^{\mu}$ & $d^{\mu}$ (\AA) & $p^{\mu}$ \\
		\midrule
		Rh--Al & 4 & 2.49 & 1.27 & Rh--Ga & 4 & 2.52 & 0.43 & Rh--In & 4 & 2.71 & 1.04 \\
		Rh--Al & 4 & 2.54 & 1.31 & Rh--Ga & 4 & 2.53 & 0.58 & Rh--In & 4 & 2.67 & 1.08 \\
		Rh--Al & 2 & 2.59 & 0.85 & Rh--Ga & 2 & 2.54 & 0.31 & Rh--In & 2 & 2.63 & 0.64 \\
		\bottomrule
	\end{tabular}

	\subsection{Optical properties }
		
	The optical properties of a material are determined by how its charge carriers interact with incident photons or electromagnetic waves. These properties are vital for applications such as optical coatings, reflectors, absorbers, and optoelectronic devices. Particularly, the response to visible light is significant for technologies like display systems and optical data storage. Energy-dependent optical features also provide insights into the electronic structure, including band structure, impurity levels, excitons, and lattice vibrations~\cite{c43}. A key parameter in optical spectroscopy is the complex dielectric function, $\varepsilon(\omega)$, which can be obtained from electronic band structure calculations:
	\begin{equation}
		\varepsilon(\omega) = \varepsilon_1(\omega) + i\varepsilon_2(\omega)\tag{17}
	\end{equation}
	The real part, $\varepsilon_1(\omega)$, describes polarization and wave propagation, while the imaginary part, $\varepsilon_2(\omega)$, relates to absorption. Both intra-band and inter-band transitions contribute to $\varepsilon(\omega)$, reflecting the material's optical and electronic properties~\cite{c44}.
	
	From $\varepsilon(\omega)$, various optical properties such as optical conductivity, absorption, refractive index, reflectivity, extinction coefficient, and energy-loss function can be derived. For LaRh$_2$X$_2$ (X = Al, Ga, In), these optical functions were calculated up to 50~eV photon energy for [100] polarization, as shown in Fig.~11 and Fig.~12.
	\subsubsection{Real and Imaginary part of the dielectric function }
	Fig.~11(a) delineates the spectral response of the real component of the dielectric function, $\varepsilon_1(\omega)$, for the investigated compounds, beginning at the zero-frequency threshold. Initially, $\varepsilon_1(\omega)$ ascends to its peak magnitude before exhibiting a progressive decline into negative values. At lower photon energies, $\varepsilon_1(\omega)$ demonstrates a resurgence, achieving maximum values of 0.86 for LaRh$_2$Al$_2$, 1.1 for LaRh$_2$Ga$_2$, and 0.28 for LaRh$_2$In$_2$ at an excitation energy of 5.25~eV, followed by a gradual descent into the negative regime~\cite{c45}. The transition of $\varepsilon_1(\omega)$ into negative values within specific energy intervals signifies the onset of complete electromagnetic wave reflection. These reflection domains are identified within the spectral ranges [$-4.6$, $-1.72$~eV at 2.6--8.2~eV] for LaRh$_2$Al$_2$, [$-3.42$, $-2.24$~eV at 2.56--7.87~eV] for LaRh$_2$Ga$_2$, and [$-4.53$, $-2.22$~eV at 2.08--6.96~eV] for LaRh$_2$In$_2$, marking their strong reflective properties in these energy regions. Conversely, the imaginary component of the dielectric function, $\varepsilon_2(\omega)$, encapsulates the optical absorption characteristics of the material, which are governed by transitions between occupied and unoccupied electronic states, derived through momentum matrix elements \cite{c47}.
	
	\begin{equation}
		\varepsilon_2 = \frac{2e^2\pi}{\Omega\varepsilon_0} \sum_{k,v,c} \left| \langle \Psi_k^c | \hat{u} \cdot \vec{r} | \Psi_k^v \rangle \right|^2 \delta(E_k^c - E_k^v - E)
		\tag{22}
	\end{equation}
	
	Where, $\varepsilon$ represents the frequency of incident radiation, $e$ denotes the elementary electronic charge, $\Omega$, $\vec{r}$ is the electron radius vector, $\hat{u}$ is the electron radius vector, and $\hat{u}$ defines the polarization.
	vector of the impinging electric field. Additionally, $\Psi_k^v$ and $\Psi_k^c$ represent the wave functions associated with the valence and conduction bands at a given wave vector $k$, respectively. The imaginary dielectric function, $\varepsilon_2(\omega)$, serves as a pivotal metric for evaluating light-matter interactions and has been rigorously analyzed for LaRh$_2$Al$_2$, LaRh$_2$Ga$_2$, and LaRh$_2$In$_2$, as depicted in Fig.~11(b). Each compound exhibits distinct absorption peaks at specific energy levels, characterizing their unique optical response. LaRh$_2$Al$_2$ attains a peak absorption of 53.8 at 0.51~eV, whereas LaRh$_2$Ga$_2$ reaches a maximum of 47.95 at 0.59~eV. In contrast, LaRh$_2$In$_2$ demonstrates superior absorption efficiency within the infrared regime, exhibiting the highest absorption peak of 64.7 at 0.41~eV, underscoring its enhanced capability to absorb light relative to the other materials.
	\subsubsection{Optical conductivity $\sigma(\omega)$}
	
	The number of free electrons in a material influences its optical conductivity. Photoconductivity refers to the ability of free charge carriers to conduct across a specific range of photon energies.
	
	\begin{equation}
		\sigma(\omega) = \frac{\omega}{4\pi} \left( \varepsilon_2(\omega) \right)
		\tag{23}
	\end{equation}
	
	As the number of free electrons increases, the optical conductivity also rises. The optical conductivity is denoted by $\sigma(\omega)$ and is measured in units of $(\Omega\,\text{m})^{-1}$. The following equation gives the mathematical expression for optical conductivity~\cite{c49}.
	
	The above-mentioned demonstrates that the optical conductivity and optical absorption are directly related. Fig.~11(c) displays the optical conductivity graph for LaRh$_2$X$_2$ (X = Al, Ga, and In). The optical spectra curves reveal that both materials exhibit a wide range of optical conductivity in the spectral region of 0--30~eV. The highest optical conductivity for LaRh$_2$Al$_2$, LaRh$_2$Ga$_2$, and LaRh$_2$In$_2$ was seen as 5.34 at 1.2~eV, 5.35 at 6.9~eV, and 5.46 at 1~eV, respectively. The photoconductivity  begin with zero photon energy that means there are no bandgap as calculated from band structure and total density of states.
	
	\begin{figure}[ht]
		\centering
		\includegraphics[width=0.9\textwidth]{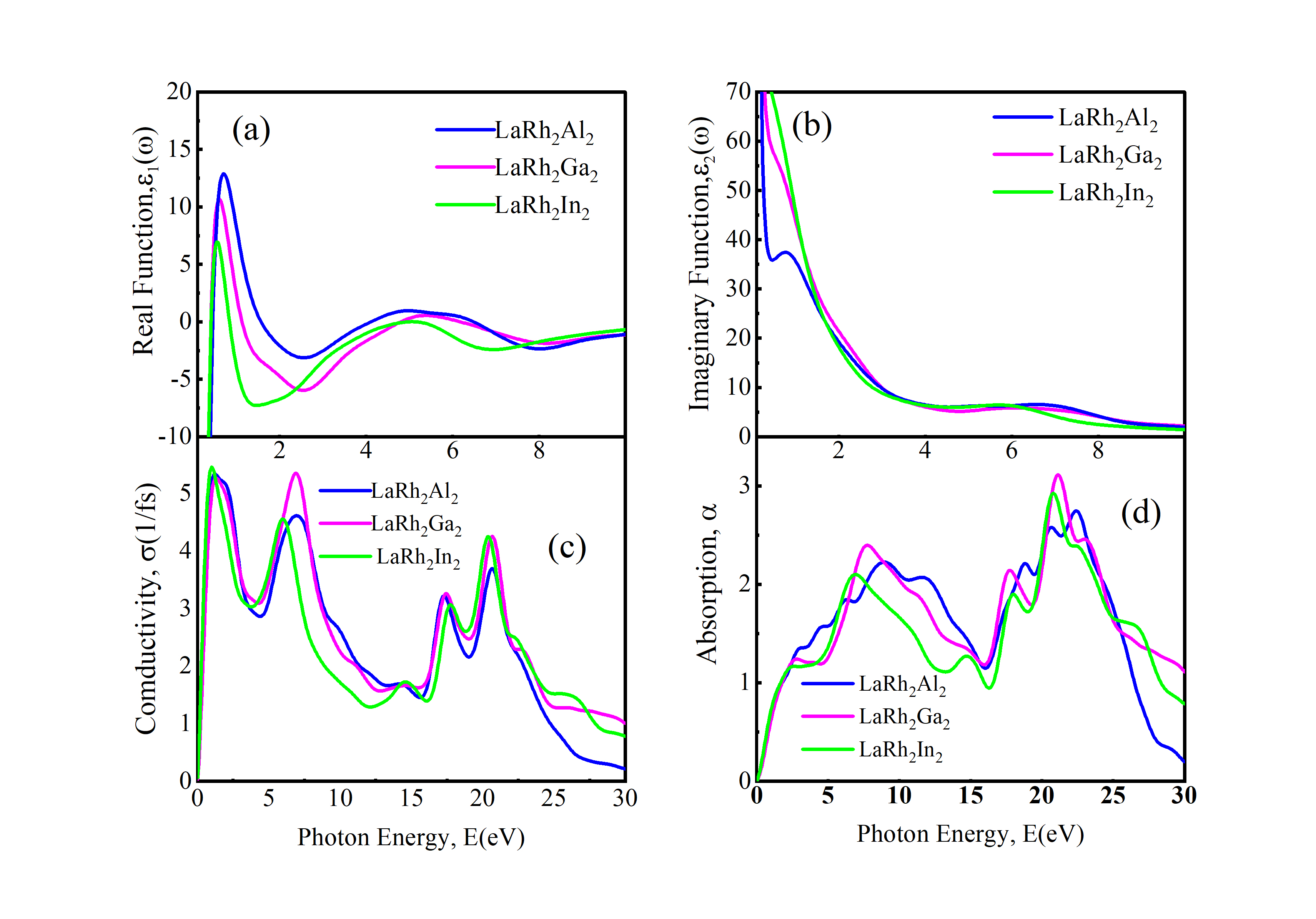}
		\caption{\label{dd7}Variation of the \textbf{real} and \textbf{imaginary} parts of the dielectric function, optical \textbf{conductivity} $\sigma$, and \textbf{absorption coefficient} $\alpha$ for (a) LaRh$_2$Al$_2$, (b) LaRh$_2$Ga$_2$, and (c) LaRh$_2$In$_2$ as a function of incident photon energy.
		}
	\end{figure}
	\subsubsection{Optical absorption coefficient}
	
	The optical absorption coefficient, $\alpha(\omega)$, measures how much light of a specific wavelength is absorbed by a material. It provides insights into the material's electrical properties, such as whether it is metallic, semiconducting, or insulating. $\alpha(\omega)$ can be determined using the real and imaginary parts of the dielectric function. The relation connecting these components to $\alpha(\omega)$ enables a deeper understanding of the material's optical behavior:
	
	\begin{equation}
		I(\omega) = \left[ \sqrt{ \varepsilon_1^2(\omega) + \varepsilon_2^2(\omega) - \varepsilon_1(\omega) } \right]^{\frac{1}{2}}
		\tag{24}
	\end{equation}
	
	The absorption in a material is influenced by its band gap and the molecular structure of the crystal. This occurs because the absorption values are frequency-dependent. Fig.~12(d) shows the absorption coefficient curves for LaRh$_2$X$_2$ (X = Al, Ga, and In). The curves show that optical absorption starts at zero photon energy, confirming the metallic nature of the compounds. This behavior underscores the unique optical properties of these materials.Fig.~12(d) shows that the absorption coefficient of LaRh$_2$X$_2$ is significant in the photon energy range of 0--30~eV. The maximum optical absorption values for LaRh$_2$Al$_2$, LaRh$_2$Ga$_2$, and LaRh$_2$In$_2$ .are 2.75, 3.11, and 2.92, occurring at photon energies of 22.4, 21.13, and 20.8 eV, respectively. These variations in absorption peaks arise from different inter-band optical transitions. This behavior aligns with the characteristics of the imaginary part of the dielectric constant. The distinct peaks highlight the influence of electronic structure on optical properties. The relationship between the transitions and dielectric behavior underscores the materials complex optical response. These findings emphasize the role of band structure in determining absorption features.
	
	\begin{figure}[ht]
		\centering
		\includegraphics[width=0.9\textwidth]{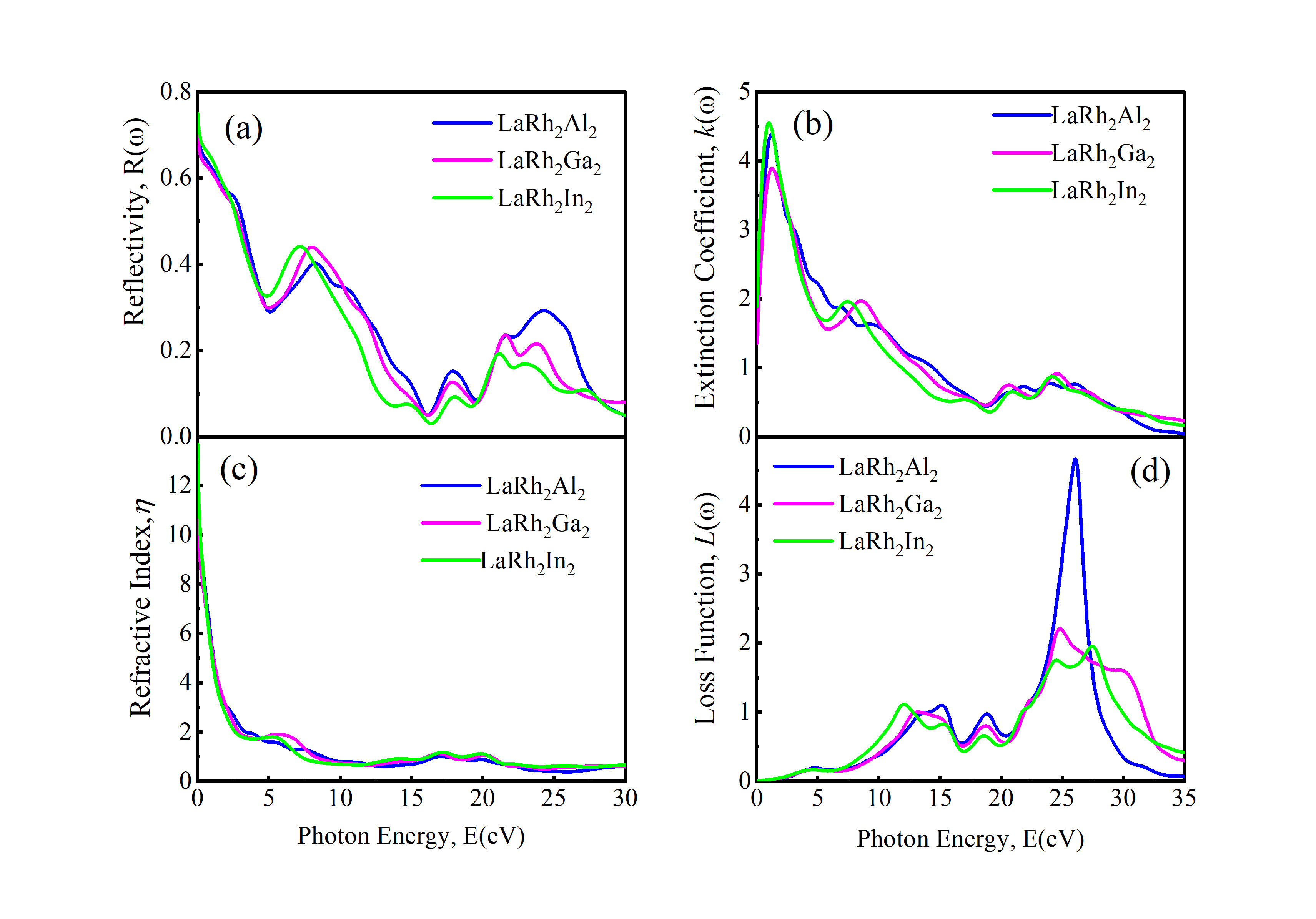}
		\caption{\label{dd8}Reflectivity $R(\omega)$, extinction coefficient $k(\omega)$, refractive index $\eta(\omega)$, and loss function $L(\omega)$ as a function of incident photon energy for (a) LaRh$_2$Al$_2$, (b) LaRh$_2$Ga$_2$, and (c) LaRh$_2$In$_2$.
		}
	\end{figure}
	\subsubsection{Reflectivity $R(\omega)$}
	
	Reflectivity measures how the power of light reflected from a material interface. It provides key insights into the materials surface and optical properties. The reflectivity can be calculated using a specific mathematical relation:
	
	\begin{equation}
		R(\omega) = \frac{(1-n)^2 + k^2}{(1+n)^2 + k^2}
		\tag{25}
	\end{equation}
	
	where $n$ is the real component and $k$ is its imaginary component of the refractive index. The optical reflectivity of the materials investigated in this study is shown in Fig.~12(a). Reflectivity starts at 0.72 for LaRh$_2$Al$_2$, 0.70 for LaRh$_2$Ga$_2$, and 0.75 for LaRh$_2$In$_2$, with the maximum reflectivity occurring at zero photon energy. Based on this data, it can be concluded that LaRh$_2$In$_2$ exhibits better reflectivity compared to LaRh$_2$Ga$_2$ and LaRh$_2$Al$_2$.
	
	\subsubsection{Extinction coefficient $k(\omega)$}
	The amount of absorption loss that occurs when an electromagnetic wave, like as light, travels through a material is described by the extinction coefficient $k(\omega)$, which may be computed using a mathematical expression. The evolution of the extinction coefficient as a function of an incident photon is depicted in Fig.~12(b).
	
	\begin{equation}
		k(\omega) = \left[ \frac{-\varepsilon_1(\omega)}{2} + \sqrt{ \frac{ \varepsilon_1^2(\omega) + \varepsilon_2^2(\omega) }{2} } \right]^{\frac{1}{2}}
		\tag{26}
	\end{equation}
	
	The maximum value of the extinction coefficient observed on the spectra is 4.55 at 0.85~eV for LaRh$_2$In$_2$, 3.9 at 1.06 for LaRh$_2$Ga$_2$ and 4.38 at 1.02 for LaRh$_2$Al$_2$. From the discussion of extinction coefficient for LaRh$_2$In$_2$ superconductor have greater attenuation of light as it passes through a medium rather than LaRh$_2$Al$_2$ and LaRh$_2$Ga$_2$.
	
	\subsubsection{Refractive index $\eta(\omega)$}
	
	The refractive index $\eta(\omega)$ is directly connected to the real dielectric function and describes the speed of light propagation at the material density obtained by mathematical expression.
	
	\begin{equation}
		\eta(\omega) = \left[ \frac{ \varepsilon_1(\omega) }{2 } + \sqrt{ \frac{ \varepsilon_1^2(\omega) + \varepsilon_2^2(\omega) }{2} } \right]^{\frac{1}{2}}
		\tag{27}
	\end{equation}
	
	Fig.~12(c) shows that the refractive indices are slight similar to $\varepsilon_1(\omega)$, in particular, the location of the peak. The values of static refractive indices $\eta(0)$ are 12.28 for LaRh$_2$Al$_2$, 11.5 for LaRh$_2$Ga$_2$ and 13.67 for LaRh$_2$In$_2$. The slight higher refractive index at $\eta(0)$, LaRh$_2$In$_2$ superconductor have greater properties such as bending of light, total internal reflection and enhance optical density rather than LaRh$_2$Al$_2$ and LaRh$_2$Ga$_2$\cite{c50}.
	
	\subsubsection{Loss function $L(\omega)$}
	The energy loss function, $L(\omega)$, quantifies the dissipative energy per unit area as a high-velocity electron traverses a material, primarily due to polarization effects, which reflect the material’s intrinsic electronic response. The interplay between absorption, reflection, and energy dissipation within a material is fundamentally interconnected, shaping its overall optical characteristics. Plasma resonance, arising from collective charge excitations, manifests as the most pronounced peak in $L(\omega)$ and is directly associated with the material’s intrinsic plasma frequency, $\omega_p$. The trailing edges of a material’s reflectance spectrum are intrinsically linked to the peak structure in its corresponding loss function, elucidating the relationship between optical and plasmonic behaviors. Mathematically the expression can be written as follows:
	
	\begin{equation}
		\omega_p^2 = \frac{4\pi N e^2}{m^*}
		\tag{28}
	\end{equation}
	
	where $N$ signifies the carrier concentration, $m^*$ represents the effective mass, and $e$ denotes the elementary charge. Spectroscopic investigations reveal that the bulk plasma frequencies occur at 26.06~eV for LaRh$_2$Al$_2$, 24.82~eV for LaRh$_2$Ga$_2$, and 27.43~eV for LaRh$_2$In$_2$, indicating variations in their electronic environments. Notably, the highest plasma frequency in LaRh$_2$In$_2$ suggests a relatively higher charge carrier density or reduced effective mass, contributing to its distinct electronic characteristics.
	\subsection{Superconducting Properties}
	
	The experimentally discovered LaRh$_2$Ga$_2$ exhibits superconductivity at low temperatures, driven by fundamental mechanisms. In conventional superconductors, the transition temperature is influenced by electron-phonon interactions and their relationship with material properties~\cite{c51}.
	
	\begin{equation}
		T_c = \frac{\theta_D}{1.45} \exp \left\{ \frac{ -1.04\,(1+\lambda) }{ \lambda - \mu^*(1+0.62\,\lambda) } \right\}
		\tag{29}
	\end{equation}
	
	Where, $\lambda$ is the electron-phonon coupling constant and $\mu^*$ is the repulsive Coulomb pseudopotential. The Coulomb pseudopotential can also be obtained from the formula as follows:
	
	\begin{equation}
		\mu^* = 0.26\, \frac{N(E_F)}{1 + N(E_F)}
		\tag{30}
	\end{equation}
	Where, $N(E_F)$ refers to total density of state at the Fermi level.
	
	This approach becomes increasingly complex for crystal systems with many atoms per unit cell, such as LaRh$_2$Ga$_2$. The intricacies of such structures make the application of this method challenging, rendering it unfeasible to proceed with this technique. An alternative approach involves using an equation to determine the electron-phonon coupling constant. This method offers a different way to analyze the coupling mechanism~\cite{c52}.
	
	\begin{equation}
		\lambda = \frac{3\gamma}{2\pi^2 k_b^2 N(E_F)} - 1
		\tag{31}
	\end{equation}
	
	In this case, experimental data for the electronic specific heat coefficient ($\gamma$) are unavailable. Additionally, theoretical calculations provide a lower value for $\lambda$. Consequently, we are left with only an indirect approach to estimate $\lambda$. This involves utilizing McMillan’s equation, as outlined below.
	
	\begin{equation}
		\lambda_{\mathrm{el-ph}} = 
		\frac{
			1.04 + \mu^* \ln \left( \frac{\theta_D}{1.45\,T_c} \right)
		}{
			\left(1 - 0.62\,\mu^*\right) \ln \left( \frac{\theta_D}{1.45\,T_c} \right) - 1.04
		}
		\tag{32}
	\end{equation}
	
	Although it is somewhat discretionary, the repulsive Coulomb potential $\mu^*$ is usually selected as 0.13. The electron-phonon coupling constant $\lambda_{\mathrm{el-ph}}$ is determined to be 0.597 for LaRh$_2$Al$_2$, 0.62 for LaRh$_2$Ga$_2$, 0.761 for LaRh$_2$In$_2$ by using McMillan's equation with $T_c = 3.7$~K~\cite{c22} and Debye temperature of 282.11~K for LaRh$_2$Al$_2$, 248.15~K for LaRh$_2$Ga$_2$ and 133.40~K for LaRh$_2$In$_2$, respectively appears to have a moderate electron-phonon coupling. This indicates that the material behaves as a moderately coupled BCS superconductor. The results align with expectations for such superconducting systems. 
	
	\section{Conclusions}
	
	In the present study, comprehensive investigation is performed using the DFT-based CASTEP code to analyze the structural, vibrational, electronic, thermophysical, optical, and superconducting characteristics of LaRh$_2$X$_2$ (where X = Al, Ga, In) superconductors. The results demonstrate the material’s structural stability, supported by their negative formation energies, as well as their mechanical stability according to the Born criteria and vibrational analysis. Pugh ratio and Poisson’s ratio indicate that the studied compounds exhibit ductile characteristics. The Debye temperature, melting point, and Vickers hardness values indicate that these materials are rather soft. LaRh$_2$Ga$_2$ has the highest melting temperature, whereas LaRh$_2$In$_2$ has a lower Debye temperature and a lower thermal conductivity. The relatively low Debye temperature of LaRh$_2$In$_2$ implies its applications for thermal barrier coatings. Electronic band structure and DOS confirm the metallic nature of all compounds where Rh-4$d$ and Rh-4$p$ orbitals significantly influence their electronic properties. The bonding analysis also indicate the metallic nature along with the coexistence of ionic and covalent bonds among the constituent elements of three compounds. The Fermi surface topology elucidates the intrinsic multi-band characteristics for all three compounds. The electron charge density analysis delineates the predominantly ionic bonding affinity between La and Rh, whereas the Rh--Rh interactions imply metallic bonding. Phonon dispersion analysis substantiates the vibrational stability of LaRh$_2$Al$_2$ and LaRh$_2$Ga$_2$, whereas LaRh$_2$In$_2$ exhibits a marginal instability at the $\Gamma$ point. Optical characteristics, including absorption, dielectric constant, loss function, reflectivity, and refractive index, have been meticulously calculated and extensively analyzed where the derived refractive index points the potential suitability of these superconductors for high-density optical data storage applications. Furthermore, the optical absorption demonstrates pronounced efficacy in the high-energy ultraviolet spectrum for promising applications as sensor, detectors, aerospace, and defence. Furthermore, the computed superconducting electron-phonon coupling constant ($\lambda = 0.56$) for LaRh$_2$Ga$_2$ signifies its classification as a weakly coupled, low-$T_c$ superconductor.

	\bibliographystyle{unsrt}
	\bibliography{Rifat.bib}
\end{document}